\documentclass{aa} 
\usepackage{txfonts}
\usepackage{longtable}
\usepackage{rotating}
\usepackage{natbib}
\usepackage{graphicx}
\usepackage{graphics}
\usepackage{psfrag}
\usepackage{amssymb}
\bibpunct{(}{)}{;}{a}{}{,}
\def\Teff{\ensuremath{T_{\mathrm{eff}}}}
\def\logg{\ensuremath{\log g}}

\def\vmic{$\upsilon_{\mathrm{mic}}$}
\def\vmac{$\upsilon_{\mathrm{macro}}$}
\def\vsini{\ensuremath{{\upsilon_{\mathrm{eq}}}\sin i}}
\def\kms{km\,s$^{-1}$}
\def\ms{m\,s$^{-1}$}

\def\espa{ESPaDOnS}

\def\logl{\ensuremath{\log L/L_{\odot}}}

\def\loglxlbol{\ensuremath{\log L_{\mathrm{X}}/L_{\mathrm{bol}}}}

\def\M{\ensuremath{M_{\odot}}}
\def\R{\ensuremath{R_{\odot}}}
\def\vald{{\sc vald}}
\def\synth{{\sc synth3}}

\def\bz{$\langle$B$_z\rangle$}
\def\cd{c\,d$^{-1}$}
\begin{document} 
\title{B fields in OB stars (BOB): on the detection of weak magnetic fields in the two early B-type stars $\beta$\,CMa and $\epsilon$\,CMa\thanks{Based on observations made with ESO Telescopes at the La Silla Paranal Observatory under programme ID\,088.A-9003(A) and ID\,191.D-0255(D,F).}}
\subtitle{Possible lack of a ``magnetic desert'' in massive stars} 
\author{L. Fossati\inst{1}    \and
        N. Castro\inst{1}     \and
	T. Morel\inst{2}      \and
	N. Langer\inst{1}     \and
	M. Briquet\inst{2}    \fnmsep\thanks{F.R.S.-FNRS Postdoctoral 								Researcher, Belgium}\and
 	T. A. Carroll\inst{3}  \and
	S. Hubrig\inst{3}      \and
	M. F. Nieva\inst{4,5}  \and
	L. M. Oskinova\inst{6} \and
	N. Przybilla\inst{5}   \and
	F. R. N. Schneider\inst{1}	   \and
	M. Sch\"oller\inst{7}              \and
	S. Sim{\'o}n-D{\'{\i}}az\inst{8,9} \and
	I. Ilyin\inst{3}	           \and
	A. de Koter\inst{10,11}             \and
	A. Reisenegger\inst{12}		   \and
	H. Sana\inst{13}		   \and
	the BOB collaboration
}
%\offprints{L.~Fossati} 
\institute{
	Argelander-Institut f\"ur Astronomie der Universit\"at Bonn, Auf dem 			H\"ugel 71, 53121, Bonn, Germany\\
	\email{lfossati@astro.uni-bonn.de}
	\and
	Institut d'Astrophysique et de G\'eophysique, Universit\'e de Li\`ege, 			All\'ee du 6 Ao\^ut, B\^at. B5c, 4000 Li\`ege, Belgium
	\and
	Leibniz-Institut f\"ur Astrophysik Potsdam (AIP), An der Sternwarte 16, 		D-14482 Potsdam, Germany
	\and
	Dr. Karl Remeis-Observatory \& ECAP, Astronomical Institute, 				Friedrich-Alexander University Erlangen-Nuremberg, Sternwartstr. 7, 			96049 Bamberg, Germany
	\and
	Institut f\"ur Astro- und Teilchenphysik, Universit\"at Innsbruck, 			Technikerstr. 25/8, 6020 Innsbruck, Austria
	\and
	Institute for Physics and Astronomy, University of Potsdam, D-14476 			Potsdam, Germany
	\and
	European Southern Observatory, Karl-Schwarzschild-Str. 2, D-85748 			Garching, Germany
	\and
	Instituto de Astrof\'isica de Canarias, 38200, La Laguna, Tenerife, 			Spain
	\and
	Departamento de Astrof\'isica, Universidad de La Laguna, E38205, La 			Laguna, Tenerife, Spain
	\and
	Astronomical Institute Anton Pannekoek, University of Amsterdam, 			Kruislaan 403, 1098 SJ, Amsterdam, The Netherlands
	\and
	Instituut voor Sterrenkunde, KU Leuven, Celestijnenlaan 200D, 3001, 			Leuven, Belgium
	\and
	Instituto de Astrof\'isica, Pontificia Universidad Cat\'olica de Chile, 		Casilla 306, Santiago 22, Chile
	\and
	European Space Agency, Space Telescope Science Institute, 3700 San 			Martin Drive, Baltimore, MD 21218, USA
} 
\date{} 
\abstract
{
Only a small fraction of massive stars seem to host a measurable structured magnetic field, whose origin is still unknown and whose implications for stellar evolution still need to be assessed. Within the context of the ``B fields in OB stars (BOB)'' collaboration, we used the HARPSpol spectropolarimeter to observe the early B-type stars $\beta$\,CMa (HD\,44743; B1\,II/III) and $\epsilon$\,CMa (HD\,52089; B1.5II) in December 2013 and April 2014. For both stars, we consistently detected the signature of a weak ($<$30\,G in absolute value) longitudinal magnetic field, approximately constant with time. We determined the physical parameters of both stars and characterise their X-ray spectrum. For the $\beta$\,Cep star $\beta$\,CMa, our mode identification analysis led to determining a rotation period of 13.6\,$\pm$\,1.2\,days and of an inclination angle of the rotation axis of 57.6\,$\pm$\,1.7$^{\circ}$, with respect to the line of sight. On the basis of these measurements and assuming a dipolar field geometry, we derived a best fitting obliquity of about 22$^{\circ}$ and a dipolar magnetic field strength (B$_{\mathrm d}$) of about 100\,G (60\,$<$\,B$_{\mathrm d}<$\,230\,G within the 1$\sigma$ level), below what is typically found for other magnetic massive stars. This conclusion is strengthened further by considerations of the star's X-ray spectrum. For $\epsilon$\,CMa we could only determine a lower limit on the dipolar magnetic field strength of 13\,G. For this star, we determine that the rotation period ranges between 1.3 and 24\,days. Our results imply that both stars are expected to have a dynamical magnetosphere, so the magnetic field is not able to support a circumstellar disk. We also conclude that both stars are most likely core hydrogen burning and that they have spent more than 2/3 of their main sequence lifetime. A histogram of the distribution of the dipolar magnetic field strength for the magnetic massive stars known to date does not show the magnetic field ``desert'' observed instead for intermediate-mass stars. The biases involved in the detection of (weak) magnetic fields in massive stars with the currently available instrumentation and techniques imply that weak fields might be more common than currently observed. Our results show that, if present, even relatively weak magnetic fields are detectable in massive stars and that more observational effort is probably still needed to properly access the magnetic field incidence.
}
% context heading (optional), leave it empty if necessary CONTEXT
% aims heading (mandatory) AIMS
% methods heading (mandatory) METHODS
% results heading (mandatory) RESULTS
\titlerunning{Weak magnetic fields in two early B-type stars}
\authorrunning{L. Fossati et al.}
\maketitle
\section{Introduction}\label{introduction}
Magnetic fields are an important constituent of astrophysical plasmas, and they play a vital role in all types of stars. On the main sequence, magnetic fields are ubiquitous in low-mass stars, presumably produced by a dynamo process at the bottom of the differentially rotating convective envelope \citep{reiners2012}. As a major consequence, these stars are spun down by magnetic braking (i.e., through the coupling of their partly ionised wind with their surface magnetic field), which is also responsible for the slow rotation of the Sun.  

In intermediate-mass main sequence stars (1.5--8\,\M), internal mostly toroidal magnetic fields \citep{spruit2002} are thought to be responsible for coupling core and envelope rotation \citep{suijs2008}, while only about 10\% of them, the magnetic chemically peculiar (CP) stars (early F-, A-, and late B-type stars), show a large scale surface magnetic field \citep{donati2009}. The magnetic CP stars are also generally spun down, giving rise to a bimodal distribution of rotational velocities for main sequence stars in this mass range \citep{royer2007,zorec2012}.  

In massive stars, internal toroidal magnetic fields may also transport angular momentum \citep{heger2005}, while their subsurface convection zones may produce small scale magnetic surface spots \citep{cantiello2009,ramia2014}. Whereas these processes are thought to occur in essentially all massive main sequence stars, only about 7\% of them seem to show large scale surface magnetic fields \citep{wade2014}. In analogy to what is observed for intermediate-mass stars \citep[see][]{royer2007,zorec2012}, this might relate to the bimodal distribution of rotational velocities in early B-type stars found by \citet{dufton2013}, although the bimodality tends to vanish with earlier spectral type \citep{oscar2013,simon2014}. 

The origin of a measurable large-scale magnetic field in about 7\% of the upper main sequence stars has not yet been understood. Except for the spin-down effect \citep{ud-Doula2009}, evolutionary consequences are also practically unexplored \citep{langer2014}. Clues may come from the field strength distribution in these stars. The intermediate-mass stars show a dichotomy, with stars being either strongly magnetic, i.e. dipolar field strengths exceeding 300\,G, or essentially non-magnetic \citep{auriere2007,lignieres2009,donati2009,petit2011}. In massive stars, while the data are sparser, applying Occam's razor would suggest a similar situation. However, as we show below, this may not be true.

With the aim of characterising the magnetic field incidence and properties in slowly rotating O- and early B-type stars, we obtained high-resolution spectropolarimetric observations of galactic O- and B-type stars in the frame of an ESO Large Programme titled ``B fields in OB stars'' (BOB) \citep{hubrig2014,morel2014}. In this context, we present here the detection of a magnetic field in the two very bright early B-type stars $\beta$\,CMa and $\epsilon$\,CMa. We find that at least $\beta$\,CMa hosts a weak magnetic field, while for $\epsilon$\,CMa the actual dipolar magnetic field strength cannot be determined yet, though we detected a weak longitudinal magnetic field. We describe our stellar parameters determination in the next section.
Section~\ref{observations} describes the observations collected for the magnetic field detection and the method adopted for their analysis. Section~\ref{results} presents the results of the magnetic field search, which are discussed in Sect.~\ref{discussion}. Conclusions are drawn in Sect.~\ref{conclusion}.
\section{Determining stellar parameters}\label{parameters}
For the atmospheric parameter and chemical abundance determination, we used both FEROS (ESO Program ID\,088.A-9003(A)) and HARPS (ESO Program ID\,191.D-0255(D,F)) spectra, because of the larger wavelength coverage of the former and the higher spectral resolution of the latter. The larger wavelength coverage of the FEROS spectra, in comparison to that of HARPS, allows one to consider an extended set of strategic spectral lines for the analysis, while the higher spectral resolution of the HARPS spectra allows one to determine more accurately the macroscopic broadening parameters. The data used for the spectroscopic analysis were obtained in December 2011 using the FEROS spectrograph, attached to the MPG/ESO 2.2\,m telescope at La Silla \citep[$R$\,=\,48\,000;][]{kaufer1999}. The spectra, collected by adopting an exposure time of 15 and 10 seconds, have a peak signal-to-noise ratio (S/N) of about 700 for $\beta$\,CMa
 and 550 for $\epsilon$\,CMa. The HARPS spectra are described in Sect.~\ref{observations}

The quantitative analyses were performed using the methodology and tools described by \citet{nieva2012}. This employs line-profile fits of synthetic to observed spectra using $\chi^2$ minimisation, aiming at a simultaneous reproduction of ionisation equilibria of \ion{He}{i/ii} (when available) and various metals, as well as the Balmer lines in an iterative approach. Thus, atmospheric parameters (effective temperature \Teff, surface gravity \logg, microturbulence \vmic, macroturbulence \vmac, projected rotational velocity \vsini), and elemental abundances are derived. The results of the analysis are given in Table~\ref{tab:params}. The models rely on hybrid non-LTE line-formation computations \citep{nieva2007}, based on hydrostatic {\sc Atlas9} LTE model atmospheres \citep{kurucz1996}, non-LTE level populations, and synthetic spectra computed with updated versions of {\sc Detail/Surface} \citep{giddings1981,butler1985}. 

Our atmospheric parameters and CNO abundances are consistent (to within the mutual error bars) with the results of \citet{morel2008}, except for the microturbulence velocities, which we find to be systematically lower. The atmosphere of $\beta$\,CMa shows a normal nitrogen abundance relative to standard values in early B-type stars \citep{nieva2012}, while $\epsilon$\,CMa is N-enriched.

 As a further check we determined the atmospheric parameters using the stellar atmosphere code {\sc fastwind} \citep[Fast Analysis of STellar atmospheres with WINDs;][]{santo1997,puls2005} and the technique described in \citet{castro2011} optimised with a genetic algorithm. In this way we obtained atmospheric parameters in excellent ($<$1$\sigma$) agreement with those listed in Table~\ref{tab:params}. 

In addition to the results of the quantitative analyses, Table~\ref{tab:params} presents the information on spectral type, $V$-band magnitude, {\sc hipparcos} distance \citep[$d_{\mathrm{HIP}}$;][]{vanleeuwen2007}, and luminosity. The last was derived from the stars' magnitude, {\sc hipparcos} distance, and bolometric correction by \citet{flower1996}, assuming no extinction. The use of different bolometric corrections \citep[for example by][]{nieva2013} leads to luminosity values within 1$\sigma$ of the adopted ones.

Both stars present narrow spectral lines, therefore we used the higher resolution HARPS spectra presented in Sect.~\ref{observations} to refine the \vsini\ and \vmac\ values obtained from the FEROS spectra. We applied the tool {\sc iacob-broad} \citep{simon2014} on the \ion{Si}{iii} 4567\,\AA\ line observed in each HARPS spectrum to estimate both \vsini\ and \vmac. Similar to what was found by \citet{aerts2014}, the analysis of spectra obtained at different epochs provides somewhat different broadening values. The situation is more critical for $\beta$\,CMa than for $\epsilon$\,CMa. The broadening parameters obtained for both stars are summarised in Table~\ref{tab:params}. Using the Fourier transform (FT) method, for $\beta$\,CMa from the available 10 HARPS spectra, we obtained \vsini\ values ranging between 13.2 and 33.6\,\kms\ with an average and standard deviation of 20.3$\pm$7.1\,\kms, which is in good agreement with what is derived from mode identification (see Sect.~\ref{astero}), while we obtained \vmac\ values ranging between 36.3 and 45.7\,\kms, with an average and standard deviation of 41.2$\pm$4.2\,\kms. Since \vsini\ is not supposed to vary with time, the large \vsini\ variations have to be attributed to the inaccurate representation of macroturbulence and pulsational broadening, as also concluded by \citet{aerts2014}. For $\epsilon$\,CMa we found from the available
eight HARPS spectra that both \vsini\ and \vmac\ are constant within the uncertainties: \vsini\,=\,21.2$\pm$2.2\,\kms\ and \vmac\,=\,46.7$\pm$2.0\,\kms. We repeated the analysis using the goodness-of-fit method obtaining results comparable to those gathered from the FT method.

We determined the stellar parameters mass $M$, radius $R$, and age $\tau$ on the basis of two different sets of evolutionary tracks by \citet{georgy2013} and \citet{brott2011} (see Table~\ref{tab:params}). We constrained the stars' $M$, $R$, and $\tau$ on the basis of the tracks by \citet{georgy2013} using the \Teff\ and \logg\ values derived from the spectroscopic analysis. For determining the stars' $M$, $R$, and $\tau$ on the basis of the Milky Way stellar evolution models of \citet{brott2011}, we used the {\sc bonnsai} code\footnote{The {\sc bonnsai} web-service is available at {\tt www.astro.uni-bonn.de/stars/bonnsai}.} \citep{schneider2013,schneider2014} and the constraints given by \logl, \Teff, \logg, and \vsini simultaneously. {\sc bonnsai} computes the posterior probability distribution of stellar model parameters given a set of observational data (\Teff, \logg, \logl, and \vsini\ in this case) using Bayes' theorem. The stellar parameters derived from the two sets of stellar evolution models agree within one sigma, except for the age of $\beta$\,CMa where the agreement is at 2$\sigma$. Both stars match the predicted nuclear path in the N/C vs. N/O diagram perfectly \citep{przybilla2010,maeder2014}. In the following, we always adopt the stellar parameters obtained using the evolution tracks by \citet{brott2011}.

As discussed below, certainly $\beta$\,CMa, but most likely also $\epsilon$\,CMa, are evolved core hydrogen-burning stars. This is in line with the spectral type and luminosity class determination given in Table~\ref{tab:params} which was derived using the relationships discussed by \citet{nieva2013}\footnote{While the spectral class could be clearly determined, the luminosity class was inferred by slight extrapolations from these relationships.} for high-resolution and high-S/N spectra. Our classification of $\beta$\,CMa agrees with the one originally given by \citet{lesh1968}. For $\epsilon$\,CMa we recommend adopting our refined classification based on high-quality data, instead of the classification given in SIMBAD. Both stars have never been reported to be in a binary system.
%--------------------------------------------------------------------
\begin{table}[h!]
\caption[]{Stellar parameters determined for $\beta$\,CMa and $\epsilon$\,CMa.}
\label{tab:params}
\begin{center}                     
\begin{tabular}{lcc}
\hline
\hline
 & $\beta$\,CMa & $\epsilon$\,CMa \\
\hline
Sp.~type              & B1\,II/III    & B1.5\,II      \\
$V$ [mag]             &  1.97         &  1.50         \\
$d_\mathrm{HIP}$ [pc] & 151$\pm$5     & 124$\pm$2     \\
\logl                 & 4.41$\pm$0.06 & 4.35$\pm$0.05 \\
\loglxlbol            & $-$7.5        & $-$7.6        \\
\hline   
\Teff\  [K]     & 24700$\pm$300  & 22500$\pm$300  \\
\logg  [cgs]    &  3.78$\pm$0.08 &  3.40$\pm$0.08 \\
\vmic\  [\kms]  &     8$\pm$1    &     8$\pm$1    \\
\vsini\  [\kms] &  20.3$\pm$7.1  &  21.2$\pm$2.2  \\
\vmac\  [\kms]  &  41.2$\pm$4.2  &  46.7$\pm$2.0  \\
%\vr\  [\kms]    &    34          &  26.6          \\
\hline
$\log n$(C)     & 8.32$\pm$0.07  & 8.30$\pm$0.07  \\
$\log n$(N)     & 7.75$\pm$0.09  & 8.16$\pm$0.07  \\
$\log n$(O)     & 8.73$\pm$0.11  & 8.70$\pm$0.12  \\
\hline
 & \multicolumn{2}{c}{G13} \\
M [\M]          & $12.0^{+0.3}_{-0.7}$ & $13.1^{+1.0}_{-0.9}$ \\
R [\R]          & $7.4^{+0.8}_{-0.9}$  & $12.0^{+1.7}_{-1.5}$ \\
$\tau$ [Myr]    & $13.8^{+2.1}_{-0.6}$ & $13.9^{+1.6}_{-1.4}$ \\
\hline
 & \multicolumn{2}{c}{B11} \\
M [\M]          & $12.6^{+0.4}_{-0.5}$ & 12.0$\pm$0.4         \\
R [\R]          & $8.2^{+0.6}_{-0.5}$  & $10.1^{+0.7}_{-0.5}$ \\
$\tau$ [Myr]    & 12.2$\pm$0.5         & $14.6^{+0.7}_{-0.6}$\\
\hline
\end{tabular}
\end{center}                     
\tablefoot{Uncertainties are 1$\sigma$-values. The \vsini\ value given for $\beta$\,CMa is that obtained from the analysis of the \ion{Si}{iii} 4567\,\AA\ line with the FT method. The last six lines list the stellar parameters mass, radius, and age obtained adopting the evolutionary models by \citet{georgy2013} -- G13 and \citet{brott2011} -- B11.}
\end{table}

%--------------------------------------------------------------------

%
\subsection{$\beta$\,CMa}\label{betaCMa_parameters}
The star $\beta$\,CMa (HD\,44743) is known to be a $\beta$\,Cep pulsator
with a primary pulsation period of 6.03096\,$\pm$\,0.00001\,hours \citep{vanleeuwen1997,shob2006}. Using EUVE observations, \citet{cassinelli1996} showed that $\beta$\,CMa has an extreme ultraviolet excess. They suggested that this may be related to the presence of heated regions near the stellar surface owing to pulsations or back-warming by the shocked wind. Making use of photometric and spectroscopic observations, \citet{mazumdar2006} performed mode identification and fitting of the pulsation frequencies obtaining the stellar parameters (mass, radius, and age) and a rotational velocity of 31\,$\pm$\,5\,\kms\  (implying a rotation period of 18.6\,$\pm$\,3.3\,days). In particular, the asteroseismic mass and radius are systematically greater than our derived values, in line with other such comparisons \citep[e.g.,][]{briquet2011}.

\begin{figure}[]
\includegraphics[width=85mm,clip]{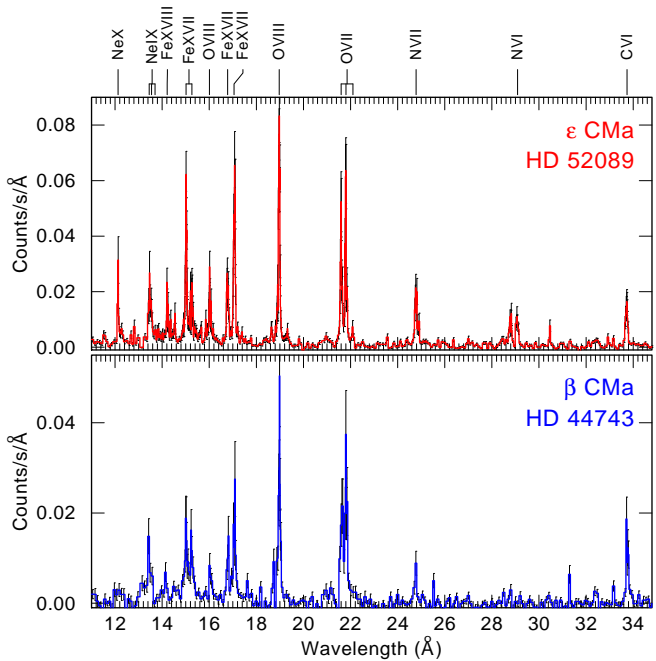}
\caption{High-resolution X-ray spectra of $\epsilon$\,CMa (top panel) and $\beta$\,CMa (bottom panel) obtained with the RGS spectrograph on board the {\em XMM-Newton} X-ray observatory. The error bars correspond to 3$\sigma$ uncertainties. The major transitions are indicated at the top.}
\label{fig:xrays}
\end{figure}

Using X-ray ROSAT observations, \citet{drew1994} derived a mass loss rate of 6($\pm$2)$\times$10$^{-9}$\,M$_\odot$\,yr$^{-1}$. We retrieved the most recent X-ray observations of $\beta$\,CMa from the
{\em XMM-Newton} archive. The observations, obtained 6 March 2008 with an exposure time of 20\,ks (observation identifier 0503500101), were reduced using recent calibrations. The X-ray flux in the 0.3--2.8\,keV band is 1.17\,($\pm$\,0.01)$\times10^{-12}$\,erg\,s$^{-1}$\,cm$^{-2}$. To estimate the
X-ray luminosity, we corrected the flux for the interstellar absorption. Within the 90\% confidence level, we estimated the X-ray luminosity of $\beta$\,CMa to be in the range 3.2--3.3$\times10^{30}$\,erg\,s$^{-1}$ (\loglxlbol\,$\approx$\,$-$7.5), in agreement with what is given by \citet{ignace2013}. This compares well with the X-ray luminosities of other $\beta$\,Cep-type variables \citep{oskinova2011}. \citet{ignace2013} find that the X-ray spectrum of $\beta$\,CMa is somewhat softer than for the magnetic $\tau$\,Sco-analog stars. The high-resolution {\em XMM-Newton} RGS spectrum of $\beta$\,CMa is shown in the lower panel of Fig.~\ref{fig:xrays}. From this spectrum we determined the ratio of fluxes between the forbidden and intercombination lines. In hot stars, this so-called f/i-line ratio indicates the proximity of the X-ray emitting plasma to the stellar surface \citep{blumen1972}. For $\beta$\,CMa, the f/i ratio for the \ion{O}{VII} line is $\approx$\,0.08, indicating that the hot plasma is formed relatively close to stellar surface. To measure the temperatures of the X-ray emitting plasma, we fitted the high- and low-resolution spectra (RGS and EPIC-PN) simultaneously. The spectra are described well by a three-temperature, optically thin plasma with temperatures of $\approx$\,1\,MK, $\approx$\,3.5\,MK, and $\approx$\,8\,MK. (The hottest plasma component is evident in the EPIC-PN data.) The weighted average emission temperature is 2.5\,MK (0.2\,keV), which is somewhat lower than what we found for $\epsilon$\,CMa (see Sect.~\ref{epsilonCMa_parameters}).

\citet{hubrig2006,hubrig2009} used observations conducted with the FORS1 low-resolution spectropolarimeter of the ESO/VLT to look for the presence of a large-scale stellar magnetic field, but the observations led to non-detections, with an upper limit on \bz\ of about 150\,G, considering a detection threshold of 5$\sigma$. A similar result was obtained by \citet{silvester2009} on the basis of \espa\ high-resolution spectropolarimetric observations: they derived an average longitudinal magnetic field value of \bz\,=\,$-$31\,$\pm$\,13\,G.
%--------------------------------------------------------------------
%--------------------------------------------------------------------

%
\subsection{$\epsilon$\,CMa}\label{epsilonCMa_parameters}
\citet{vallerga1995} suggested that $\epsilon$\,CMa (HD\,52089) is the major contributor of hydrogen-ionising photons in the solar neighbourhood. Although located in the $\beta$\,Cep instability strip, this star is not known to pulsate. We analysed archival {\em XMM-Newton} observations of $\epsilon$\,CMa obtained on 19 March 2001 using an exposure time of 45\,ks (observation identifier 0069750101). The X-ray flux in the 0.3--2.8\,keV band is 1.31\,($\pm$\,0.05)$\times10^{-12}$\,erg\,s$^{-1}$. To estimate the X-ray luminostity, we corrected the flux for the interstellar absorption. Within the 90\% confidence level, we estimated the X-ray luminosity of $\epsilon$\,CMa to be in the range 2.3--2.7$\times10^{30}$\,erg\,s$^{-1}$ (\loglxlbol\,$\approx$\,$-$7.6). The high-resolution X-ray spectrum is shown in the top panel of Fig.~\ref{fig:xrays}. The spectrum is described well by a three-temperature, optically thin plasma with temperatures $\approx$\,1\,MK, $\approx$\,3\,MK, and $\approx$\,6\.MK. The weighted average emission temperature is 3.6\,MK (0.31\,keV), which is quite similar to that of other magnetic B-type stars \citep{oskinova2011}. From fitting the RGS spectra we found a nitrogen overabundance, in agreement with the optical analysis. The width of the X-ray emission lines indicates that the hot plasma is not expanding faster than 400\,\kms. Interestingly, the f/i-line ratio for $\epsilon$\,CMa is 0.09, which is very similar to what is found for other B-type stars with different magnetic field strengths \citep[e.g.,][]{oskinova2014}.

Hamann et al. (priv. communication) performed non-LTE analysis of multi-wavelength spectra (X-ray, EUV, UV, optical, and IR) of $\epsilon$\,CMa using the stellar atmosphere code PoWR \citep{graefener2002,hamann2004}. From fitting the UV resonance lines of C, N, and Si, they derived a mass-loss rate of $1\times10^{-9}$\,M$_\odot$\,yr$^{-1}$ and a terminal wind velocity of $v_\infty$\,=\,700\,\kms. The terminal wind velocity is in good agreement with the previous determinations \citep[e.g.,][]{snow1976}, while the mass-loss rate is about a factor of ten lower than crudely estimated by \citet{drew1994}.

\citet{hubrig2009} report a magnetic field detection for this star on the basis of average longitudinal magnetic field values (\bz) of $-$200\,$\pm$\,48\,G and $-$129\,$\pm$\,34\,G obtained from FORS1 observations conducted in November 2006 and August 2007, respectively. The re-analysis of the FORS1 data conducted by \citet{bagnulo2012} showed instead that the first detection (2006 data) might be spurious (\bz\,=\,$-$127\,$\pm$\,60\,G), while they obtained a 5.3$\sigma$ detection from the 2007 data (\bz\,=\,$-$196\,$\pm$\,37\,G), in agreement with the result of \citet{hubrig2009}.
\subsection{Evolutionary status}\label{hrd}
\begin{figure*}[]
\sidecaption
\includegraphics[width=129mm,clip]{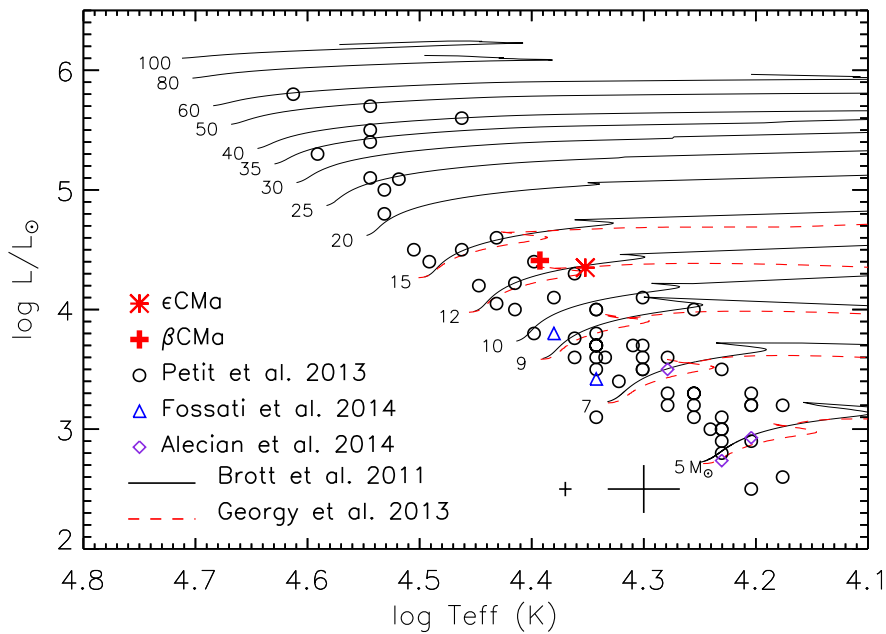}
\caption{Position of $\beta$\,CMa (cross) and $\epsilon$\,CMa (asterisk) in the Hertzsprung-Russell diagram in comparison with that of other magnetic stars. The luminosities for the magnetic stars published by \citet{alecian2014} have been computed on the basis of the stars' magnitude and distance, the latter given by \citet{alecian2014}. The evolutionary tracks for Milky Way metallicity by \citet{brott2011} and the evolutionary tracks for solar metallicity by \citet{georgy2013} are overplotted. Each track is labelled with its initial mass in units of the solar mass. The larger cross on the bottom of the plot shows the size of the median uncertainty in \logl\ and \Teff, while the smaller cross shows the size of the uncertainties in \logl\ and \Teff\ for $\beta$\,CMa and $\epsilon$\,CMa.}
\label{fig:hr}
\end{figure*}

Figure~\ref{fig:hr} shows the position of $\beta$\,CMa and $\epsilon$\,CMa in the Hertzsprung-Russell diagram (HRD), together with that of most of the magnetic massive stars published so far \citep{petit2013,fossati2014,alecian2014}. For reference, Fig.~\ref{fig:hr} also shows the evolutionary tracks for non-rotating stars by \citet{georgy2013} and \citet{brott2011}. The HRD does not include the magnetic massive star in the Trifid nebula published by \citet{hubrig2014} because the complicated nature of the system has not yet allowed it to be determined which is/are the magnetic star/s or the stars' basic parameters.

According to the evolutionary tracks of \citet{brott2011}, both $\beta$\,CMa and $\epsilon$\,CMa are still in their main-sequence evolutionary phase. Comparing with the evolutionary tracks by \citet{georgy2013} gives the same result for $\beta$\,CMa, but $\epsilon$\,CMa might already be a post-main sequence object. However, since this would put $\epsilon$\,CMa into an extremely rapid evolutionary state and since a slightly higher temperature or a slightly larger overshooting parameter than used by \citet{georgy2013} would recover it as a main sequence star, we consider it more likely that $\epsilon$\,CMa is still undergoing core hydrogen burning. Both stars have then spent more than two-thirds of their main sequence time. 

Figure~\ref{fig:hr} covers massive magnetic stars and the upper end of the intermediate-mass magnetic stars. The dividing line between massive and intermediate-mass stars is usually drawn at about 8\,\M\ on the basis of their expected fate. Figure~\ref{fig:hr} suggests instead that there is no dividing line in terms of magnetic properties (e.g., distribution across the HR diagram). %--------------------------------------------------------------------
%--------------------------------------------------------------------
%
\section{Observations and analysis method}\label{observations}
We observed $\beta$\,CMa and $\epsilon$\,CMa with the HARPSpol polarimeter \citep{snik2011,piskunov2011} feeding the HARPS spectrograph \citep{mayor2003} attached to the ESO 3.6-m telescope in La\,Silla, Chile. The observations, covering the 3780--6910\,\AA\ wavelength range with a spectral resolution $R$=115\,000, were obtained using the circular polarisation analyser. We observed each star with one or more sequences of four sub-exposures obtained by rotating the quarter-wave retarder plate by 90$^\circ$ after each exposure, i.e. 45$^\circ$, 135$^\circ$, 225$^\circ$, and 315$^\circ$. The observations were performed in December 2013 and April 2014. The observing journal is given in Table~\ref{tab:obs_log}. The exposure times were tuned on the basis of the seeing, which turned out to be very variable for the 2013 run. The resulting S/Ns of the Stokes $I$ spectra are listed in Table~\ref{tab:Bfield}. The spectra of $\beta$\,CMa obtained on 24 December 2013 were saturated and therefore not included in the analysis.
%
%--------------------------------------------------------------------
\begin{table}[h!]
\caption[]{Journal of the HARPS observations.}
\label{tab:obs_log}
\begin{center}                     
\begin{tabular}{llll}
\hline
\hline
Star  	& Date & HJD$-$  & Exp time \\
name 	&      & 2\,456\,000 & [s]      \\
\hline
$\beta$\,CMa    & 23/12/2013     & 650.8615 & 4$\times$60  \\
                & 24/12/2013     & 651.5584 & 4$\times$250$^a$ \\
                & 26/12/2013     & 653.7335 & 4$\times$150$^b$ \\
                & 27/12/2013     & 654.8186 & 4$\times$75  \\
                & 28/12/2013     & 655.8087 & 4$\times$130$^c$ \\
                & 21/04/2014     & 769.4650 & 4$\times$40  \\
                & 21/04/2014     & 769.4701 & 4$\times$40  \\
                & 21/04/2014     & 769.4741 & 4$\times$40  \\
                & 21/04/2014     & 769.4777 & 4$\times$40  \\
                & 21/04/2014     & 769.4813 & 4$\times$40  \\
                & 21/04/2014     & 769.4848 & 4$\times$40  \\
$\epsilon$\,CMa & 23/12/2013     & 650.8659 & 4$\times$40  \\
                & 24/12/2013     & 651.7392 & 4$\times$35  \\
                & 27/12/2013     & 654.8645 & 4$\times$55  \\
                & 21/04/2014     & 769.4907 & 4$\times$45  \\
                & 21/04/2014     & 769.4943 & 4$\times$45  \\
                & 21/04/2014     & 769.4980 & 4$\times$45  \\
                & 21/04/2014     & 769.5016 & 4$\times$45  \\
                & 21/04/2014     & 769.5052 & 4$\times$45  \\

\hline
\end{tabular}
\end{center}                     
\tablefoot{The date, given in column two, corresponds to the start of the night of observation in format dd/mm/yyyy. The heliocentric Julian date (HJD) is that of the middle of the observation. (a) Frames saturated. (b) The first frame was taken with an exposure time of 200\,s. (c) The given exposure time of 130\,s is an average of the actual exposure times which were 75, 140, 160, and 150\,s.}
\end{table}

%--------------------------------------------------------------------
%

We reduced and calibrated the data with the {\sc reduce} package \citep{piskunov2002}, obtaining one-dimensional spectra that were combined using the ``ratio'' method in the way described by \citet{bagnulo2009}. We then re-normalised all spectra to the intensity of the continuum obtaining a spectrum of Stokes $I$ ($I/I_c$) and $V$ ($V/I_c$), plus a spectrum of the diagnostic null profile \citep[$N$ - see][]{bagnulo2009}, with the corresponding uncertainties.

To detect magnetic fields, we used the least-squares deconvolution technique \citep[LSD;][]{donati1997}, which combines line profiles (assumed to
all have the same shape) centred at the position of the individual lines given in the line mask and scaled according to the line strength and sensitivity to a magnetic field (i.e., line wavelength and Land{\'e} factor). The resulting average profiles ($I$, $V$, and $N$) were obtained by combining several lines, yielding a strong increase in S/N and therefore sensitivity to polarisation signatures. We computed the LSD profiles of Stokes $I$, $V$, and of the null profile using the methodology and the code described in \citet{kochukhov2010}. 

We prepared the line mask used by the LSD code separately for each star, adopting the stellar parameters given in Sect.~\ref{parameters}. We extracted the line parameters from the Vienna Atomic Line Database \citep[\vald;][]{vald1,vald2,vald3} and tuned the given line strength to the observed Stokes $I$ spectrum with the aid of synthetic spectra calculated with \synth\ \citep{kochukhov2007}. For each star we used all lines stronger than 10\% of the continuum (considering only natural broadening), avoiding hydrogen lines and lines in spectral regions affected by the presence of telluric features. For each star we performed the LSD analysis using two different line masks: with and without helium lines. The number of lines adopted in each line mask and for each star is listed in Table~\ref{tab:Bfield}.

We defined the magnetic field detection making use of the false alarm probability \citep[FAP;][]{donati1992}, considering a profile with FAP~$<10^{-5}$ as a definite detection (DD), $10^{-5}<$~FAP~$<10^{-3}$ as a marginal detection (MD), and FAP~$>10^{-3}$ as a non-detection (ND). To further check that the magnetic field detections are not spurious, we calculated the FAP for the null profile in the same velocity range as used for the magnetic field measurement, obtaining ND in all cases but one (see Table~\ref{tab:Bfield}). We also calculated the FAP for both Stokes $V$ and the null profile outside the range covered by the Stokes $I$ spectral line and as wide as the one used for the magnetic field detection, again obtaining ND in all cases. In addition, we checked whether both Stokes $V$ and the null profile are consistent with the expected noise properties (i.e., whether the Stokes $V$ uncertainties are consistent with the standard deviation of the null profile; whether the integral of Stokes $V$ and of the null profile in the range used for the magnetic field detection are consistent with zero). As a further check, we derived the Stokes $V$ LSD profiles using a line mask that contains either lines with a low Land{\'e} factor or lines with a high Land{\'e} factor. As expected for profiles that carry the signature of a magnetic field, the amplitude of the Stokes $V$ LSD profiles reflects the changes in the average Land{\'e} factor \citep[see, e.g.,][]{lignieres2009}.

We further analysed the HARPSpol spectra using the moment technique \citep{mathys1991,mathys1994} and the multi-line singular value decomposition (SVD) method \citep{carroll2012}. For this analysis, the reduction and wavelength calibration were performed using the standard HARPS data reduction pipeline, while the continuum normalisation was performed following \citet{hubrig2013}. The spectra were combined using the ratio method \citep{donati1997,bagnulo2009} to derive Stokes $I$ and $V$, while the null profile was obtained following \citet{ilyin2012}. The analysis of the SVD profiles led to FAPs comparable to those obtained from the LSD profiles. Using both techniques we also obtained magnetic field values that agree well with those derived from the LSD profiles. In the following, we always adopt the results obtained from the LSD profiles. For $\epsilon$\,CMa, using the moment technique we also detected the presence of crossover \citep{mathys1995a} and of a quadratic field \citep{mathys1995b} at a statistically significant level ($>3\,\sigma$).
\section{Results}\label{results}
\subsection{$\beta$\,CMa}
\begin{figure*}[]
\includegraphics[width=85mm,clip]{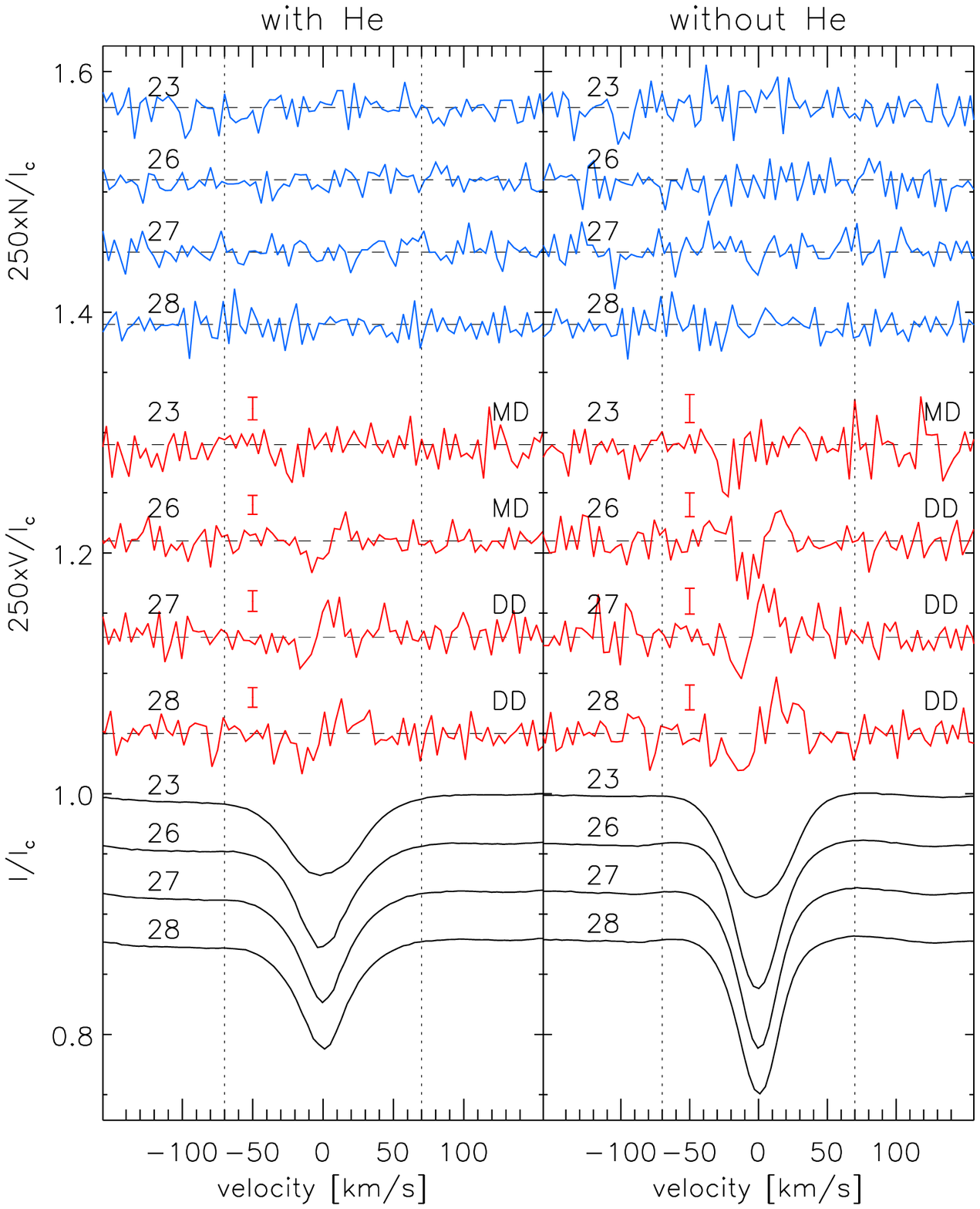}
\hspace{1cm}
\includegraphics[width=85mm,clip]{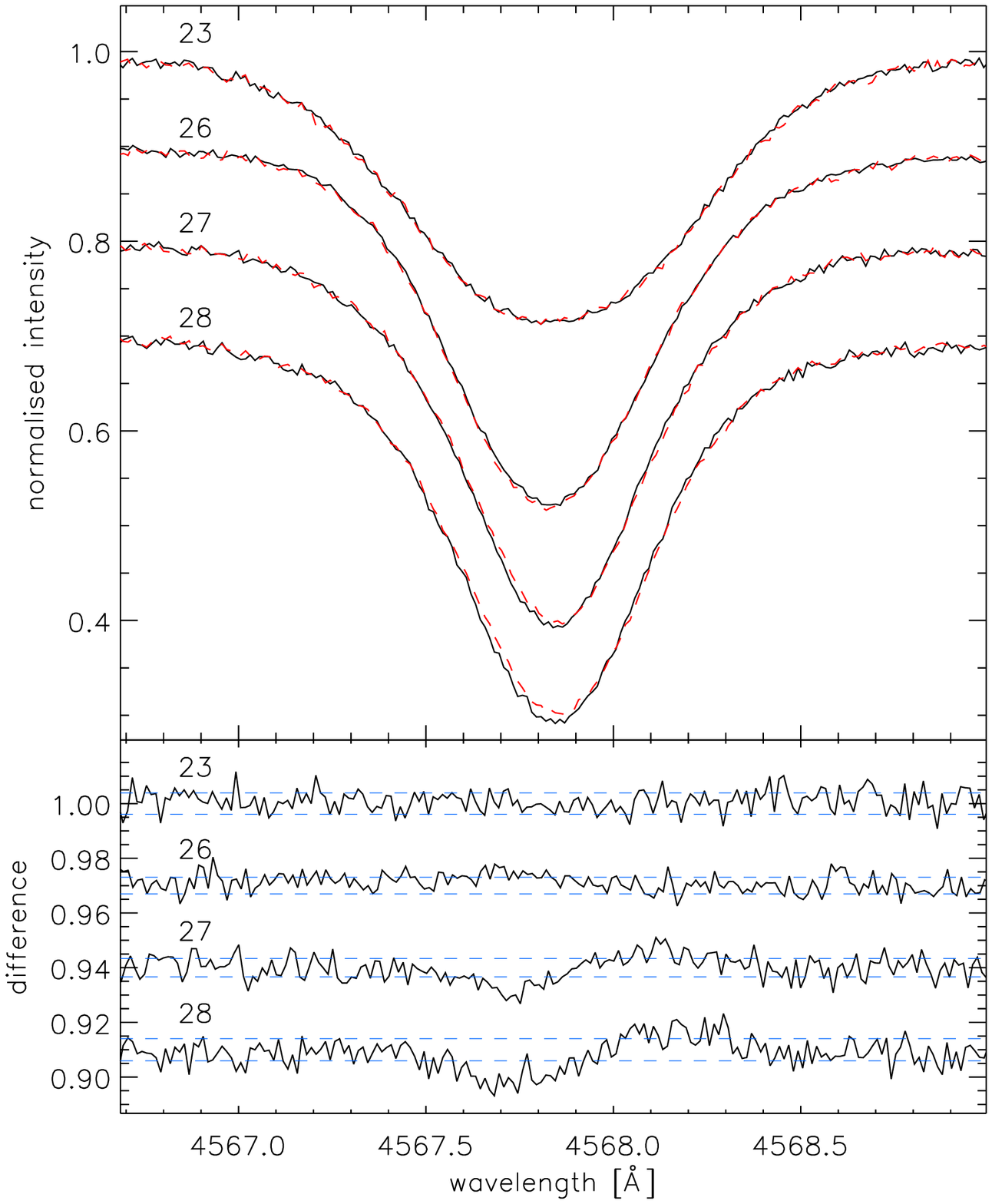}
\caption{Left plot: LSD profiles of Stokes $I$ (black solid line), $V$ (red solid line), and $N$ parameter (blue solid line) obtained for $\beta$\,CMa between 23 and 28 December 2013. The left panel shows the profiles obtained using a line mask containing He lines, while the right panel shows the profiles obtained with a line mask that does not contain He lines. The night of observation is given in the top left corner of each profile, and the FAP-based field detection is given in the top right corner of each Stokes $V$ profile. The bar at $-$50\,\kms\ shows the average uncertainty for each Stokes $V$ profile. The vertical dotted lines indicate the velocity range adopted for determining the detection probability and magnetic field value. All profiles have been rigidly shifted upwards/downwards by arbitrary values and the Stokes $V$ and $N$ profiles have been expanded 250 times. Right plot - top panel: comparison between the profiles of the \ion{Si}{iii} 4568\,\AA\ line recorded in the four nights of observation in the parallel beam with the retarder waveplate at +45$^{\circ}$ (black solid line) and +225$^{\circ}$ (red dashed line). One \kms\ corresponds to about 0.015\,\AA. Right plot - bottom panel: difference between the profiles shown in the top panel in each night of observation. For reference, the blue dashed lines indicate the standard deviation of each difference spectrum. In both panels, the number given in the top left corner of each profile indicates the night of observation in December 2013.}
\label{fig:hd44743-lsd1}
\end{figure*}

The plot on the left-hand side of Fig.~\ref{fig:hd44743-lsd1} shows the LSD profiles derived from the data obtained for $\beta$\,CMa in December 2013, while Table~\ref{tab:Bfield} gives the results gathered from their analysis. From each Stokes $V$ profile we obtained either marginal or definite detections with \bz\ values consistently below 25\,G, in absolute value.
%--------------------------------------------------------------------
%--------------------------------------------------------------------
%--------------------------------------------------------------------
\begin{table*}[]
\caption[]{Results from the LSD analysis of the HARPSpol data obtained using a line mask that includes (wHe) or excludes (woHe) helium lines.}
\label{tab:Bfield}
\begin{center}                     
\begin{tabular}{lccccccccccc}
\hline
\hline
Star \&    & Date & Mask & $<$B$_z>(V)$ & FAP~($V$) & Detection & FAP~($N$) & Detection & S/N & S/N	     & \# lines \\
month/Year &      &      & [G]          &           & $V$       &           & $N$	  & $I$ & $V_{LSD}$  &  	\\
\hline
$\beta$\,CMa& 23 & wHe  &  $-$8.4$\pm$6.8 & 5.6$\times$10$^{-4}$  & {\bf MD} & 1.0$\times$10$^{-1}$ &      ND  & 664 & 26712 & 204 \\
December    & 26 & wHe  &  $-$3.3$\pm$5.6 & 9.7$\times$10$^{-4}$  & {\bf MD} & 2.7$\times$10$^{-1}$ &      ND  & 681 & 31809 &     \\ 	  
2013        & 27 & wHe  & $-$22.4$\pm$6.1 & 1.7$\times$10$^{-8}$  & {\bf DD} & 6.3$\times$10$^{-1}$ &      ND  & 628 & 29191 &     \\ 	  
            & 28 & wHe  &  $-$5.8$\pm$5.9 & 2.3$\times$10$^{-9}$  & {\bf DD} & 1.8$\times$10$^{-4}$ & {\bf MD} & 710 & 30536 &     \\ 	  
            & 23 & woHe &  $-$9.4$\pm$8.4 & 1.1$\times$10$^{-4}$  & {\bf MD} & 4.8$\times$10$^{-2}$ &      ND  & 664 & 21436 & 161 \\
            & 26 & woHe &  $-$1.3$\pm$6.9 & 8.0$\times$10$^{-8}$  & {\bf DD} & 1.5$\times$10$^{-2}$ &      ND  & 681 & 25355 &     \\ 	  
            & 27 & woHe & $-$19.4$\pm$7.5 & 5.1$\times$10$^{-9}$  & {\bf DD} & 5.3$\times$10$^{-1}$ &      ND  & 628 & 23319 &     \\ 	  
            & 28 & woHe & $-$16.1$\pm$7.2 & 2.1$\times$10$^{-11}$ & {\bf DD} & 1.1$\times$10$^{-1}$ &      ND  & 710 & 24270 &     \\ 	  
\hline
$\beta$\,CMa& 21 & wHe  & $-$29.0$\pm$8.0 & 4.2$\times$10$^{-4}$  & {\bf MD} & 2.0$\times$10$^{-1}$ &      ND  & 554 & 21488 & 204 \\
April       & 21 & wHe  & $-$27.8$\pm$7.9 & 3.2$\times$10$^{-3}$  &      ND  & 4.9$\times$10$^{-1}$ &      ND  & 607 & 22821 &     \\
2014        & 21 & wHe  & $-$28.8$\pm$7.7 & 9.2$\times$10$^{-4}$  & {\bf MD} & 8.3$\times$10$^{-1}$ &      ND  & 589 & 23610 &     \\
            & 21 & wHe  & $-$21.9$\pm$7.5 & 1.3$\times$10$^{-5}$  & {\bf MD} & 3.7$\times$10$^{-1}$ &      ND  & 611 & 23930 &     \\
            & 21 & wHe  & $-$24.0$\pm$7.9 & 1.1$\times$10$^{-2}$  &      ND  & 8.4$\times$10$^{-1}$ &      ND  & 596 & 22413 &     \\
            & 21 & wHe  & $-$26.4$\pm$7.9 & 3.6$\times$10$^{-1}$  &      ND  & 1.7$\times$10$^{-2}$ &      ND  & 572 & 22674 &     \\
Average     & 21 & wHe  & $-$26.0$\pm$3.2 & 2.7$\times$10$^{-14}$ & {\bf DD} & 9.6$\times$10$^{-1}$ &      ND  &  -  & 55602 &     \\
            & 21 & woHe & $-$40.4$\pm$10.1& 3.6$\times$10$^{-3}$  &      ND  & 1.3$\times$10$^{-1}$ &      ND  & 554 & 17110 & 161 \\
            & 21 & woHe & $-$15.6$\pm$9.7 & 3.4$\times$10$^{-3}$  &      ND  & 2.6$\times$10$^{-2}$ &      ND  & 607 & 18202 &     \\
            & 21 & woHe & $-$27.3$\pm$9.4 & 1.4$\times$10$^{-3}$  &      ND  & 3.4$\times$10$^{-1}$ &      ND  & 589 & 18863 &     \\
            & 21 & woHe & $-$18.8$\pm$9.2 & 3.6$\times$10$^{-8}$  & {\bf DD} & 6.9$\times$10$^{-1}$ &      ND  & 611 & 19124 &     \\
            & 21 & woHe & $-$22.9$\pm$9.7 & 1.4$\times$10$^{-5}$  & {\bf MD} & 7.0$\times$10$^{-1}$ &      ND  & 596 & 17934 &     \\
            & 21 & woHe & $-$15.8$\pm$9.6 & 7.3$\times$10$^{-3}$  &      ND  & 5.0$\times$10$^{-2}$ &      ND  & 572 & 18112 &     \\
Average     & 21 & woHe & $-$23.2$\pm$3.9 & 3.3$\times$10$^{-16}$ & {\bf DD} & 9.7$\times$10$^{-1}$ &      ND  &  -  & 44703 &     \\
%\hline
%HD\,44743   &    & wHe  &$-$27.1$\pm$14.3 & 5.8$\times$10$^{-3}$  &      ND  & 2.1$\times$10$^{-1}$ &      ND  & 918 & 16901 & 204 \\
%\espa\      &    & woHe & $-$8.0$\pm$12.9 & 4.2$\times$10$^{-4}$  & {\bf MD} & 8.1$\times$10$^{-1}$ &      ND  &     & 13575 & 161 \\
\hline
$\epsilon$\,CMa& 23 & wHe  & $-$18.4$\pm$4.7 & 5.3$\times$10$^{-5}$  & {\bf MD} & 4.4$\times$10$^{-1}$ &      ND  & 634 & 30733 & 153 \\
December       & 24 & wHe  &  $-$1.5$\pm$5.9 & 5.6$\times$10$^{-2}$  &      ND  & 6.3$\times$10$^{-2}$ &      ND  & 590 & 24345 &     \\ 	  
2013           & 27 & wHe  &     0.0$\pm$4.3 & 5.4$\times$10$^{-4}$  & {\bf MD} & 5.7$\times$10$^{-1}$ &      ND  & 679 & 33806 &     \\ 	  
               & 23 & woHe & $-$11.0$\pm$5.8 & 1.5$\times$10$^{-6}$  & {\bf DD} & 4.1$\times$10$^{-1}$ &      ND  & 634 & 21268 & 119 \\
               & 24 & woHe &     9.3$\pm$7.3 & 3.7$\times$10$^{-6}$  & {\bf DD} & 2.9$\times$10$^{-1}$ &      ND  & 590 & 16789 &     \\ 	  
               & 27 & woHe &     4.6$\pm$5.3 & 3.1$\times$10$^{-11}$ & {\bf DD} & 3.7$\times$10$^{-1}$ &      ND  & 679 & 23398 &     \\ 	  
\hline
$\epsilon$\,CMa& 21 & wHe  &  $-$6.6$\pm$4.6 & 2.4$\times$10$^{-2}$  &      ND  & 4.6$\times$10$^{-2}$ &      ND  & 648 & 31485 & 153 \\
April          & 21 & wHe  &  $-$8.8$\pm$4.6 & 2.8$\times$10$^{-2}$  &      ND  & 3.7$\times$10$^{-1}$ &      ND  & 631 & 31261 &     \\
2014           & 21 & wHe  &  $-$1.4$\pm$4.6 & 1.3$\times$10$^{-1}$  &      ND  & 6.2$\times$10$^{-3}$ &      ND  & 646 & 31766 &     \\
               & 21 & wHe  &  $-$8.6$\pm$4.6 & 1.6$\times$10$^{-1}$  &      ND  & 7.9$\times$10$^{-1}$ &      ND  & 645 & 31630 &     \\
               & 21 & wHe  &     3.0$\pm$4.6 & 4.7$\times$10$^{-4}$  & {\bf MD} & 2.8$\times$10$^{-2}$ &      ND  & 606 & 31571 &     \\
Average        & 21 & wHe  &  $-$4.5$\pm$2.1 & 9.4$\times$10$^{-6}$  & {\bf DD} & 1.5$\times$10$^{-1}$ &      ND  &  -  & 69580 &     \\
               & 21 & woHe &  $-$9.3$\pm$5.6 & 7.8$\times$10$^{-5}$  & {\bf MD} & 1.4$\times$10$^{-1}$ &      ND  & 648 & 21705 & 119 \\
               & 21 & woHe &     3.7$\pm$5.7 & 1.4$\times$10$^{-3}$  &      ND  & 5.5$\times$10$^{-1}$ &      ND  & 631 & 21533 &     \\
               & 21 & woHe &     3.9$\pm$5.6 & 4.7$\times$10$^{-6}$  & {\bf DD} & 5.2$\times$10$^{-3}$ &      ND  & 646 & 21831 &     \\
               & 21 & woHe &    15.2$\pm$5.6 & 7.8$\times$10$^{-3}$  &      ND  & 6.3$\times$10$^{-2}$ &      ND  & 645 & 21726 &     \\
               & 21 & woHe &     4.0$\pm$5.6 & 2.8$\times$10$^{-2}$  &      ND  & 3.5$\times$10$^{-1}$ &      ND  & 606 & 21756 &     \\
Average        & 21 & woHe &  $-$4.1$\pm$2.5 &         $<$10$^{-16}$ & {\bf DD} & 6.4$\times$10$^{-1}$ &      ND  &  -  & 48535 &     \\
%\hline
%HD\,52089   &    & wHe  &$-$35.7$\pm$20.2 & 1.1$\times$10$^{-1}$  &      ND  & 6.9$\times$10$^{-5}$ & {\bf MD} & 839 & 16791 & 153 \\
%\espa\      &    & woHe &   24.2$\pm$15.6 & 5.3$\times$10$^{-3}$  &      ND  & 4.9$\times$10$^{-1}$ &      ND  &     &  9267 & 119 \\

\hline
\end{tabular}
\end{center}                     
\tablefoot{The first two columns indicate the star name and the date of observation. Column three indicate the adopted line mask. Column four lists the \bz\ values obtained from each Stokes $V$ LSD profile, for which the FAP is given in column five. Column six shows the detection flag obtained for each Stokes $V$ LSD profile, where marginal (MD) and definite (DD) detections are marked in bold face (ND stays for non-detection). Column seven and eight give the FAP obtained from each LSD profile of the null spectrum and the relative detection flag, respectively. Column nine gives the S/N (per-pixel) of Stokes $I$, calculated over an 8\,\AA\ region at $\sim$4990\,\AA. Column ten lists the S/N of the LSD Stokes $V$ profile. The last column lists the number of lines used in the line mask. For $\beta$\,CMa and $\epsilon$\,CMa, respectively, we adopted a range of 140\,\kms\ (i.e., $\pm$70\,\kms\ from the line center) and 120\,\kms\ (i.e., $\pm$60\,\kms\ from the line center) for the calculation of the magnetic field.}
\end{table*}

%--------------------------------------------------------------------

Because we observed $\beta$\,CMa using rather long exposure times in relation to the stellar pulsation period (3--4\% of the $\sim$6-hour period), we checked the stability of the spectral lines along each sequence, therefore highlighting movements of the reference frame. To do this, within each night of observation, we compared the profiles of the \ion{Si}{iii} $\lambda$4568 line recorded in the parallel beam with the retarder waveplate at +45$^{\circ}$ and +225$^{\circ}$ (see the plot on the right side of Fig.~\ref{fig:hd44743-lsd1}). This test is similar to the one employed by \citet{bagnulo2013} to look for instabilities in FORS (low resolution) and HARPSpol (high resolution) spectropolarimetric data. In the absence of instrumental instabilities (very unlikely in the case of HARPS) and/or significant stellar pulsations, the two profiles should be identical within the noise. The plot on the right-hand side of Fig.~\ref{fig:hd44743-lsd1} shows a clear shift in the observed line profiles obtained on 27 and 28 December. Given the pulsating nature of the star and the adopted exposure times in comparison to the star's pulsation period, we ascribed these instabilities to pulsations. Furthermore, the observed shifts are of the order of 3--6\,\kms\  (i.e., 0.05--0.1\,\AA), which is much greater than the expected HARPS instrumental instabilities of 1--2\,\ms\ \citep{lovis2006}. We
also performed a similar check for other stars (both magnetic and non-magnetic) observed in the same nights, without finding any significant line shifts, except for the pulsating stars. This supports the conclusion that the instabilities observed for $\beta$\,CMa are due to pulsations alone. 

For high-resolution observations, these shifts do not hamper the magnetic field detection \citep{schnerr2006,neiner2012,alecian2014}. To
instead determine the impact of these variations on the obtained \bz\ values, we used Gaussian synthetic lines having parameters (e.g., depth, width, land\'e factor) identical to that of the LSD profiles obtained on 27 and 28 December 2013. Using the synthetic Gaussian profiles, we reconstructed the observations, including/excluding the shifts caused by the pulsation and an input \bz\ value of 20\,G. The magnetic field was applied to the profiles using Eq.~1 of \citet{mathys1991}. We combined the synthetic lines to extract both Stokes $I$ and $V$ and then measured the magnetic field in the same way as with the observed LSD profiles. We found that shifts of the same magnitude as those registered on the spectra gathered on 27 and 28 December 2013 have a negligible impact on the measured \bz\ value.

\begin{figure*}[]
\includegraphics[width=85mm,clip]{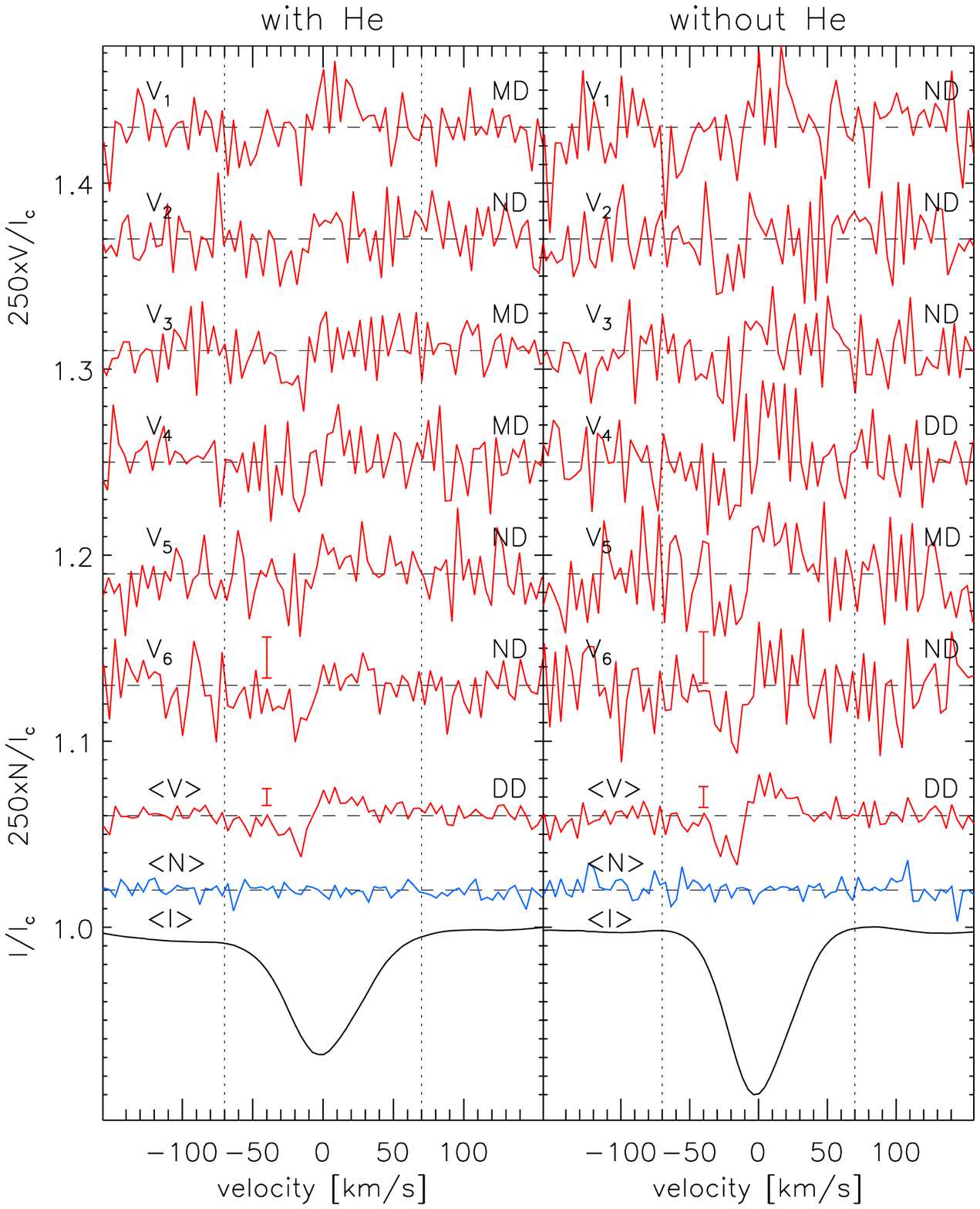}
\hspace{1cm}
\includegraphics[width=85mm,clip]{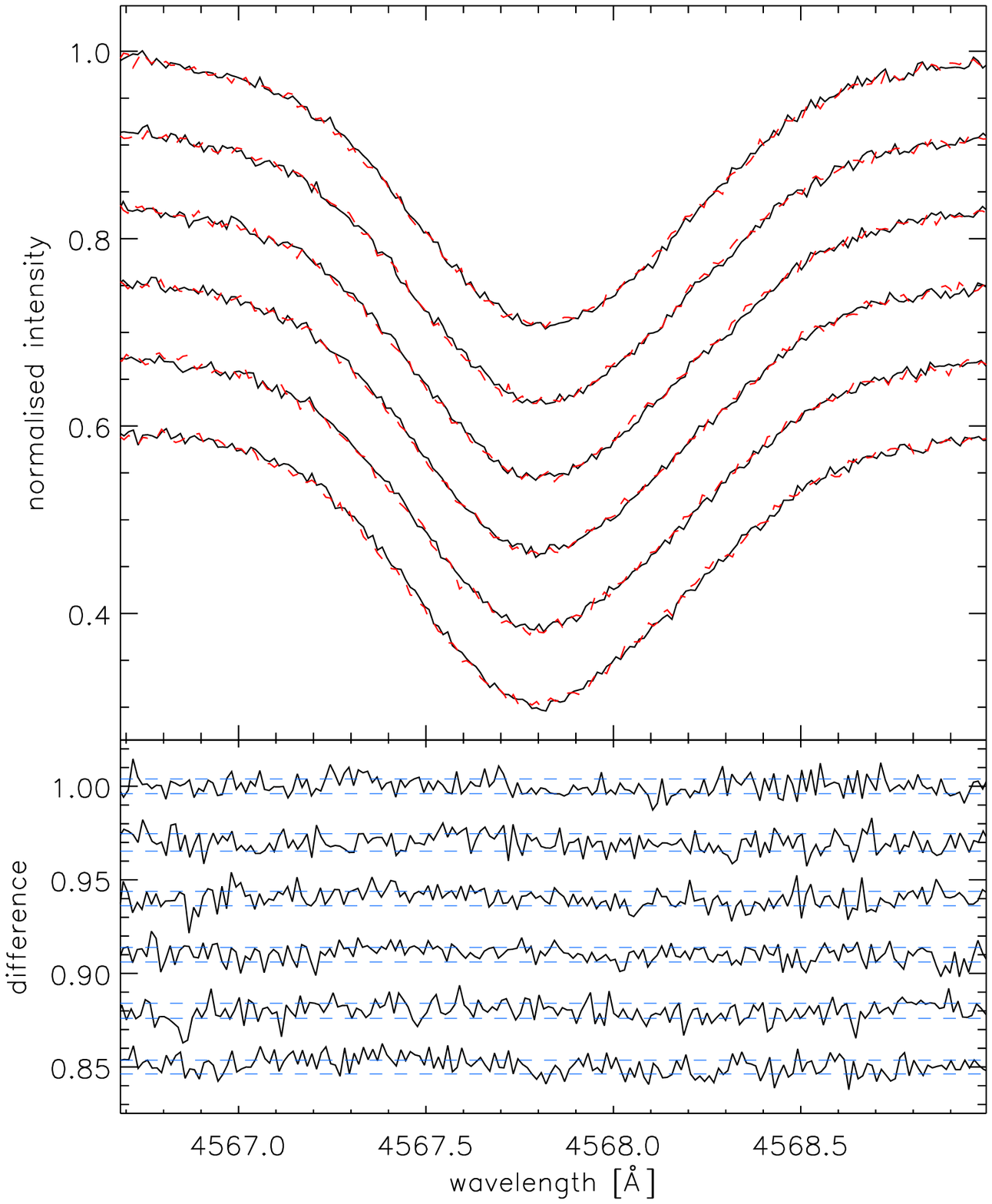}
\caption{Left plot: from bottom to top: weighted average Stokes $I$ LSD profile (black solid line), weighted average LSD profile of the diagnostic $N$ parameter (blue solid line), weighted average Stokes $V$ LSD profile (red solid line), and six Stokes $V$ LSD profiles obtained from the six single observations of $\beta$\,CMa obtained on 21 April 2014. The left panel shows the profiles obtained using a line mask containing He lines, while the right panel shows the profiles obtained with a line mask that does not contain He lines. The FAP-based field detection for Stokes $V$ (see Sect.~\ref{observations}) is given in the top right corner of each Stokes $V$ profile. The two bars at $-$40\,\kms\ show the average uncertainty of the last single Stokes $V$ profile and of the average Stokes $V$ profile. The vertical dotted lines indicate the velocity range adopted for determining the detection probability and magnetic field value. All profiles have been rigidly shifted upwards/downwards by arbitrary values, and the Stokes $V$ and $N$ profiles have been expanded 250 times. Right plot: same as the plot on the right side of Fig.~\ref{fig:hd44743-lsd1}, but for the sequence of observations obtained on 21 April 2014, in sequential order from top to bottom.}
\label{fig:hd44743-lsd2}
\end{figure*}

Nevertheless, to obtain a ``cleaner'' magnetic field detection, we re-observed the star on 21 April 2014. We obtained six identical {\it consecutive} sequences, one sequence composed by one observation at each of the four position angles (i.e., 45$^{\circ}$, 135$^{\circ}$, 225$^{\circ}$, and 315$^{\circ}$). We used LSD to analyse each single sequence and then weighted-averaged the LSD profiles in order to obtain one set of Stokes $I$, Stokes $V$, and $N$ parameter LSD profiles. By fitting a Gaussian to each Stokes $I$ LSD profile, we brought all $I$, $V$, and $N$ LSD profiles to the rest frame before averaging. Each observation was performed with a rather short exposure time (4$\times$40\,s) to make sure stellar pulsations did not affect the stability of the reference frame (see the plot on the right side of Fig~\ref{fig:hd44743-lsd2}). By averaging the single profiles we greatly increased the S/N of the LSD profiles, without saturating the HARPS CCDs. 

The plot on the left-hand side of Fig.~\ref{fig:hd44743-lsd2} shows the LSD profiles derived from the data obtained on 21 April 2014 for $\beta$\,CMa.  Table~\ref{tab:Bfield} lists the results gathered from their analysis. Given the shorter exposure times compared to the 2013 observations, the noise of the individual consecutive Stokes $V$ LSD profiles was in most cases too high to lead to a field detection. However, by averaging the profiles, we obtained extremely high S/N values that led to a solid definite detection. For $\beta$\,CMa we derived an average longitudinal magnetic field value of about $-$25\,G with an uncertainty of 3--4\,G, depending on the adopted line mask. 
%--------------------------------------------------------------------
%--------------------------------------------------------------------

The HARPSpol magnetic field detection for $\beta$\,CMa presented here contrasts with the non-detection by \citet{silvester2009}, though both analyses had been performed on high-resolution spectropolarimetric data and with the same technique (i.e., LSD). The disagreement is nevertheless only apparent: the higher resolution of HARPS, in comparison to that of \espa, allowed us to reach the S/N needed to detect the magnetic field. For a given wavelength bin, the higher resolution of HARPS allowed us to collect more photons without saturating the CCDs, and by rebinning the data using LSD, we obtained average profiles with a higher S/N. 
\subsubsection{Constraints on the stellar inclination angle and equatorial rotational velocity}\label{astero}
To characterise the magnetic field geometry and dipolar field strength (see Sect.~\ref{geometry_and_strength_hd44743}), it is first necessary to estimate the star's inclination angle and rotational velocity. Given the pulsating nature of $\beta$\,CMa, this can be done using mode identification techniques.

An intensive multisite photometric study of $\beta$\,CMa has been presented by \citet{shob2006}. They observed three pulsation frequencies and determined the degree $\ell$ of the two dominant modes. \citet{mazumdar2006} then used ground-based high resolution high S/N spectroscopic measurements to add further constraints on the $m$-values of the pulsation modes and to deduce an equatorial rotational velocity of $v_{\rm eq}\,=\,31\,\pm\,5$\,\kms. The spectroscopic mode identification was performed by means of the moment method described in \citet{briquet2003}, which could limit the range of values for the combination (\vsini, $i$), but not for \vsini or the stellar inclination angle $i$, separately.

Another mode identification method based on spectroscopy is the Fourier parameter fit (FPF) method that is implemented in the software package FAMIAS\footnote{FAMIAS was developed in the framework of the FP6 European Coordination Action HELAS -- \url{http://www.helas-eu.org/}} \citep{zima2008}. This method includes first-order effects of the Coriolis force in the modelling of the displacement field, and it was successfully applied to several $\beta$\,Cep stars, such as 12\,Lac \citep{desmet2009} and V2052\,Oph \citep{briquet2012}. As illustrated in these studies, the simultaneous fitting of a couple of non-radial modes constrains the values of $v_{\rm eq}$ and $i$ separately. Therefore, we applied it to the spectra presented in \citet{mazumdar2006} by following the same procedure as explained in \citet{briquet2012}. 

The wavenumbers ($\ell,m$) obtained by the moment method are confirmed for the two dominant modes with $f_1\,=\,3.9793$\,\cd\ and $f_2\,=\,3.9995$\,\cd, i.e., $(\ell_1,m_1)$\,=\,$(2,2)$ and $(\ell_2,m_2)$\,=\,$(0,0)$. In \citet{mazumdar2006}, the only constraint for $f_3\,=\,4.1832$\,\cd\ is $m_3>0$. The FPF method excludes $\ell_3>3$, and the best match between the observed and theoretical amplitude and phase across the line profile is obtained for ($\ell_3,m_3$)\,=\,(2,1), though one cannot rule out the solution with ($\ell_3,m_3$)\,=\,(1,1). 

\begin{figure}
\vspace{-2.4cm}
\includegraphics[width=90mm,clip]{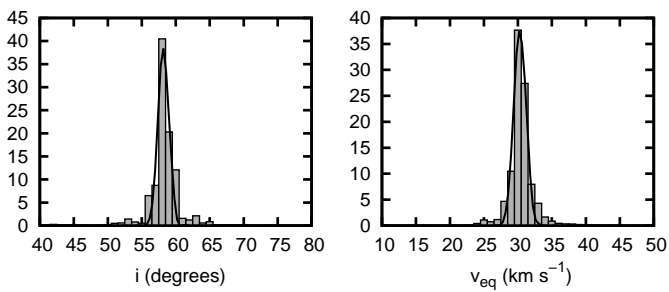}
\vspace{-6.8cm}
\caption{Histograms for the inclination and equatorial rotational velocity of $\beta$\,CMa derived from the spectroscopic mode identification performed with the FPF method assuming ($\ell_3,m_3$)\,=\,2,1.} 
\label{fig:histo}
\end{figure}

By fitting the three modes simultaneously, we derived estimates for $v_{\rm eq}$ and $i$. The solution with ($\ell_3,m_3$)\,=\,(2,1) gives $v_{\rm eq}\,=\,30.3\,\pm\,0.9$\,\kms\ and $i\,=\,58.2\,\pm\,0.8^{\circ}$. Histograms for $v_{\rm eq}$ and $i$ for this solution are shown in Fig.~\ref{fig:histo}. The solution with ($\ell_3,m_3$)\,=\,(1,1) leads to very similar values: $v_{\rm eq}\,=\,30.9\,\pm\,0.9$\,\kms\ and $i\,=\,55.3\,\pm\,1.5^{\circ}$. We point out that the moment method and the FPF method lead to fully compatible values for the equatorial rotational velocity of the star, which gives us confidence in the deduced value. Here we adopted the average of the results obtained with the FPF method on the basis of the two solutions for ($\ell_3,m_3$). From the equatorial rotational velocity $v_{\rm eq}\,=\,30.6\,\pm\,0.9$\,\kms\ we derived a rotation period of 13.6\,$\pm$\,1.2\,days. The difference between this value and the one derived by \citet{mazumdar2006} is due to the use of different values of the stellar radius. The value of the stellar radius adopted by \citet{mazumdar2006} was obtained from the best fitting seismic model, but, as discussed in detail in Sect.~\ref{geometry_and_strength_hd44743}, the seismic modelling should be redone taking into account that the star is magnetic which would most likely lead to a different best fitting model.
%--------------------------------------------------------------------
%--------------------------------------------------------------------
%
\subsubsection{Magnetic field geometry and strength}\label{geometry_and_strength_hd44743}
\begin{figure}[]
\includegraphics[width=85mm,clip]{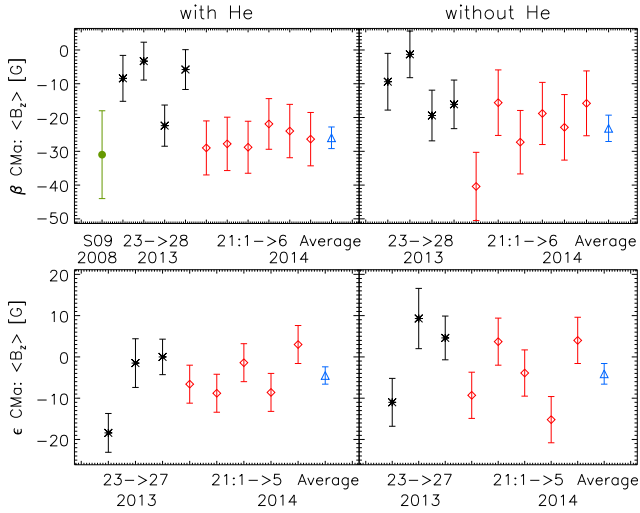}
\caption{Time series of the \bz\ values obtained for $\beta$\,CMa (top panels) and $\epsilon$\,CMa (bottom panels) using a line mask that includes (left panels) or excludes (right panels) helium lines. The black asterisks show the \bz\ values derived from the observations carried out in December 2013. The red rhombs show the \bz\ values obtained from each consecutive observation carried out in April 2014. The blue triangles indicate the \bz\ value extracted from the average LSD profiles obtained on 21 April 2014. The green dot on the left side of the top left panel shows the \bz\ value obtained by \citet{silvester2009} (S09) from the \espa\ observation of $\beta$\,CMa carried out in 2008.}
\label{fig:bz}
\end{figure}

A comparison of the Stokes $V$ LSD profiles obtained in 2013 and 2014 (about 3.5 months apart) shows that the configuration of the magnetic field facing Earth did not seem to change much over that period of time. To illustrate this, the top panels of Fig.~\ref{fig:bz} show the time series of the \bz\ values obtained for $\beta$\,CMa using the two adopted line masks (i.e., with and without helium lines). It is also interesting to notice that there is a similarity between the shape of the Stokes $V$ LSD profile obtained by \citet{silvester2009} and those presented here for $\beta$\,CMa. This is reflected by the fact that we obtained \bz\ values for $\beta$\,CMa in agreement with those of \citet{silvester2009} (see the top left panel of Fig.~\ref{fig:bz}). 
%--------------------------------------------------------------------
%--------------------------------------------------------------------

Nevertheless, variations at the level of 2$\sigma$ are indeed present, and the pulsational constraints on the rotation period and inclination angle allowed us to attempt a preliminary modelling of the star's magnetic field, despite the small number of measurements, assuming a perfect dipole \citep{preston1967,borra1979}. We performed a $\chi^2$ minimisation of a sinusoidal wave function: 
\begin{equation}
<B_{z}>(t)=A\sin(\frac{2\,\pi\,t}{P}+\phi)+ZP,
\end{equation}
where the zero point ZP, the amplitude A, the period P, and the phase $\phi$ are the variables to fit simultaneously. We constrained the period to vary only within 13.6\,$\pm$\,1.2\,days. Using the \bz\ values obtained from the mask containing He lines, we obtained the best fit for ZP=$-$16.0\,G, A=10.0\,G, P=13.77\,days, and $\phi$=92$^{\circ}$, with a $\chi^2$ of 7.263 and 6 degrees of freedom. We then determined the best fitting obliquity $\beta$ and dipolar magnetic field strength B$_{\mathrm d}$ assuming an inclination angle of 56.7$^{\circ}$ and a limb-darkening coefficient of 0.6, obtaining $\beta$=22.3$^{\circ}$ and B$_{\mathrm d}$=96.9\,G. At the 1$\sigma$ level, $\beta$ ranges between about 5 and 90$^{\circ}$, while B$_{\mathrm d}$ ranges between about 60 and 230\,G. It is important to notice that the available data points did not allow us to constrain the period better than the pulsational analysis because almost all periods within the adopted 13.6\,$\pm$\,1.2\,days range would fit the data within 1$\sigma$.

\begin{figure}[]
\includegraphics[width=85mm,clip]{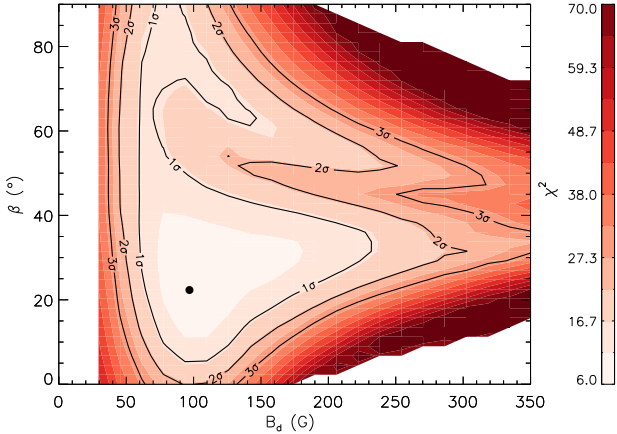}
\caption{$\chi^2$ map of dipole field strength B$_{\mathrm d}$ versus obliquity $\beta$ permitted by the longitudinal field measurements of $\beta$\,CMa, assuming an inclination angle of 56.7$^{\circ}$ and a limb darkening coefficient of 0.6. The black dot indicates the best fit.}
\label{fig:beta_Bpol}
\end{figure}

\begin{figure}[]
\includegraphics[width=85mm,clip]{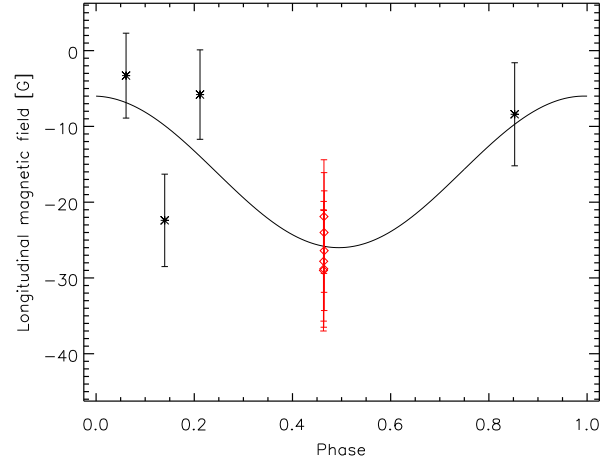}
\caption{Phase plot of the \bz\ values obtained for $\beta$\,CMa from the HARPS data and the best fitting sine wave function. The symbols are as in the top left panel of Fig.~\ref{fig:bz}.}
\label{fig:phase_plot}
\end{figure}

Figure~\ref{fig:beta_Bpol} shows the possible $\beta$-B$_{\mathrm d}$ combinations, while Fig.~\ref{fig:phase_plot} shows the phase plot of the adopted \bz\ values using ephemeris based on the best fitting set of parameters and the time of the first HARPS observation. Figure~\ref{fig:beta_Bpol} shows that at the 3$\sigma$ level, the dipolar magnetic field strength is below 300\,G. As a check, we also did the fit using the \bz\ values obtained with the mask that does not contain He lines, arriving at a very similar result: ZP=$-$19.0\,G, A=16.0\,G, P=13.44\,days, $\phi$=112$^{\circ}$, $\beta$=28.9$^{\circ}$, and B$_{\mathrm d}$=121.7\,G with a $\chi^2$ of 6.179 and 6 degrees of freedom. Even including the measurement by \citet{silvester2009} in the fit does not significantly modify the results: ZP=$-$16.5\,G, A=10.0\,G, P=13.66\,days, $\phi$=92$^{\circ}$, $\beta$=21.7$^{\circ}$, and B$_{\mathrm d}$=99.5\,G with a $\chi^2$ of 7.577 and 7 degrees of freedom. Because the spectropolarimetric observations of $\beta$\,CMa have been obtained at random times and led to rather low \bz\ values,  the presence of a dipolar magnetic field with a large amplitude (hence a large ZP in absolute value) is unlikely. Figure~\ref{fig:bz} might also be suggestive of a constant magnetic field, in which case the only possible configuration would be one in which the magnetic field axis and rotation axis are aligned, i.e., $\beta$=0$^{\circ}$. Figure~\ref{fig:beta_Bpol} shows that this possibility is excluded at the 2$\sigma$ level. Future spectropolarimetric observations will allow us to further constrain the magnetic field geometry and strength, as well as to look for the possible presence of variations in the magnetic field in phase with the pulsation.
%--------------------------------------------------------------------
%--------------------------------------------------------------------
%--------------------------------------------------------------------
%--------------------------------------------------------------------

The dipolar magnetic field strength obtained for $\beta$\,CMa is below what is typically found for other magnetic massive stars \citep[see, e.g.,][]{petit2013,alecian2014}, though it is not completely unusual \citep{donati2006,bouret2008,petit2013}. The weakness of the magnetic field is strengthened further by the results of \citet{ignace2013}. They compared the X-ray and UV properties of the magnetic stars HD\,63425 and HD\,66665, believed to be analogues of $\tau$\,Sco \citep{petitV2011}, to $\beta$\,CMa, $\xi^1$\,CMa, and $\tau$\,Sco itself, where $\beta$\,CMa was used as the non-magnetic reference star. They find that the X-ray spectrum of $\beta$\,CMa is softer than the magnetic $\tau$\,Sco-analogue stars. Following the results of \citet{ignace2013}, if $\beta$\,CMa hosted a field of the same strength as that of $\tau$\,Sco and its analogues, the differences in the X-ray spectral characteristics observed by \citet{ignace2013} would not be present. 

\citet{briquet2012} present the results of an asteroseismic analysis and modelling of the magnetic $\beta$\,Cep star V2052\,Oph (B$_{\mathrm d}\sim$400\,G). They conclude that V2052\,Oph's pulsational properties are best fit by an evolutionary model with no or mild overshooting between 0.0 and 0.15 pressure scale heights (H$_p$). \citet{briquet2012} conclude that the absence of core overshooting is probably due to the magnetic field, which inhibits mixing in the core. They also conclude that a fossil (i.e., present at least on the zero-age main-sequence; ZAMS) surface magnetic field strength of 2\,G would be enough to inhibit differential rotation and therefore rotational mixing in early B-type stars. Following \citet{briquet2012} we determined the critical magnetic field (B$_{\mathrm crit}$) for $\beta$\,CMa above which mixing might be suppressed. According to the formulation of \citet{mathis2005}, which is applicable only in the case of $\beta$\,=\,0$^{\circ}$, we found that the average magnetic field strength in the radiative zone necessary for suppressing differential rotation in $\beta$\,CMa is B$_{\mathrm crit}\sim$43\,G, corresponding to a critical surface field of about 1\,G.

Following this, we expect rigid interior rotation and no core overshooting for $\beta$\,CMa. This contradicts the results of \citet{mazumdar2006} who found instead a best fitting value of the overshooting parameter of 0.20\,$\pm$\,0.05\,H$_p$. Nevertheless, \citet{mazumdar2006} used the first overtone as radial mode, while a comparison with the stellar atmospheric parameters given in Sect.~\ref{parameters} shows that the radial mode should be taken as the fundamental mode. This clearly shows that the modelling of this star needs to account for the presence of the magnetic field. As a consequence, it might well be that a model with no overshooting would best fit the observables. For further improvement in the stellar modelling, we will need to collect more pulsation modes. These may be extracted from the light curves that will be obtained by the BRITE satellites \citep{weiss2014}.

In the context of the classification of massive stars magnetospheres presented by \citet{petit2013}, we obtained a Keplerian corotation radius of 6.8 stellar radii\footnote{We used the terminal wind velocity of 700\,\kms\ measured for $\epsilon$\,CMa. This is justified by the similarities in the stellar parameters of the two stars.} and, considering the B$_{\mathrm d}$ values obtained within the 1$\sigma$ level, an Alfv{\'e}n radius ranging between 2.1 and 3.1 stellar radii. With these results, $\beta$\,CMa has a dynamical magnetosphere, meaning that no circumstellar disk can be supported by the magnetic field. This conclusion is observationally supported by the fact that the HARPS data analysed here do not show any spectral variability (including in the H$\alpha$ line) beyond what is expected for a pulsating star. To the best of our knowledge there is also no mention in the literature of spectral variations that could be due to the presence of a disk.
\subsection{$\epsilon$\,CMa}
\begin{figure}[]
\includegraphics[width=85mm,clip]{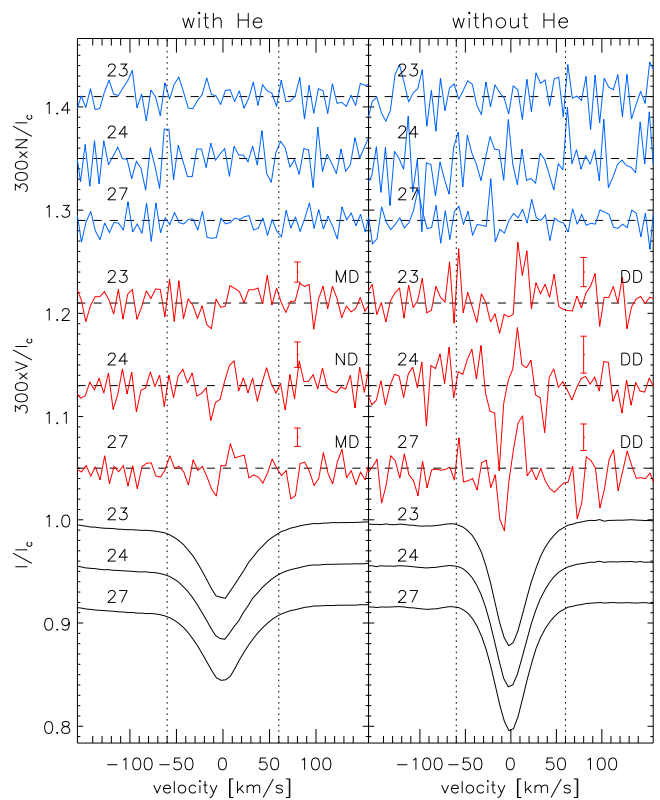}
\caption{Same as the plot on the left side of Fig.~\ref{fig:hd44743-lsd1}, but for $\epsilon$\,CMa. The Stokes $V$ and $N$ profiles have been expanded 300 times. The bar at 60\,\kms\ shows the average uncertainty for each Stokes $V$ profile.}
\label{fig:hd52089-lsd1}
\end{figure}

Figure~\ref{fig:hd52089-lsd1} shows the LSD profiles derived from the data obtained for $\epsilon$\,CMa in December 2013, while Table~\ref{tab:Bfield} gives the results gathered from their analysis. From almost every Stokes $V$ profile, we obtained either marginal or definite detections with \bz\ values consistently below 20\,G in absolute value. Also for this star we performed the same stability checks as for $\beta$\,CMa, without detecting any significant shift (see Fig.~\ref{fig:hd52089-profiles1}).
%--------------------------------------------------------------------
%--------------------------------------------------------------------
%--------------------------------------------------------------------
\onlfig{
\begin{figure}[]
\includegraphics[width=85mm,clip]{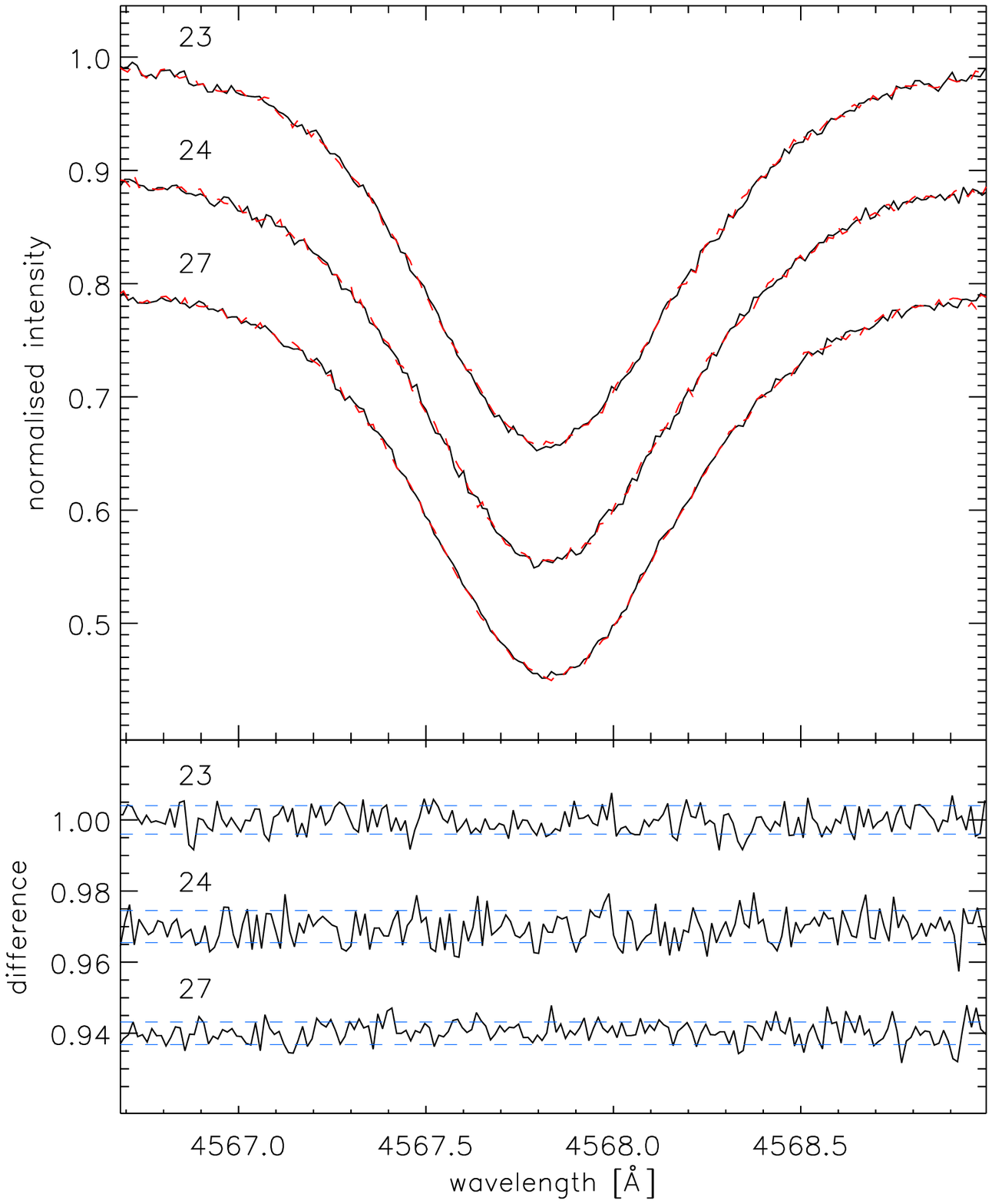}
\caption{Same as the plot on the right side of Fig.~\ref{fig:hd44743-lsd1}, but for $\epsilon$\,CMa.}
\label{fig:hd52089-profiles1}
\end{figure}
}
%--------------------------------------------------------------------

\begin{figure}[h]
\includegraphics[width=85mm,clip]{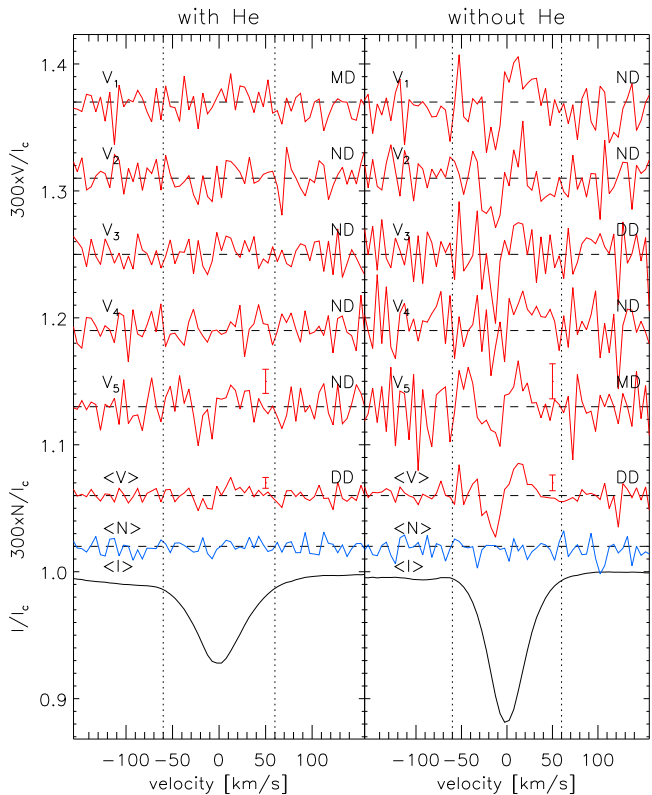}
\caption{Same as Fig.~\ref{fig:hd44743-lsd2}, but for $\epsilon$\,CMa. The Stokes $V$ and $N$ profiles have been expanded 300 times. The two bars at 50\,\kms\ show the average uncertainty for the last single Stokes $V$ profile and for the weighted average Stokes $V$ profile.}
\label{fig:hd52089-lsd2}
\end{figure}

To increase the significance of the detection we re-observed the star on 21 April 2014, using the same strategy as for $\beta$\,CMa: five identical consecutive sequences, one sequence being composed of one observation at each of the four position angles (i.e., 45$^{\circ}$, 135$^{\circ}$, 225$^{\circ}$, and 315$^{\circ}$). We then analysed the data in the same way as for $\beta$\,CMa. Figure~\ref{fig:hd52089-lsd2} shows the LSD profiles derived from the data obtained on 21April 2014, and Table~\ref{tab:Bfield} lists the results gathered from their analysis.
%--------------------------------------------------------------------
%--------------------------------------------------------------------
%--------------------------------------------------------------------
\onlfig{
\begin{figure}[]
\includegraphics[width=85mm,clip]{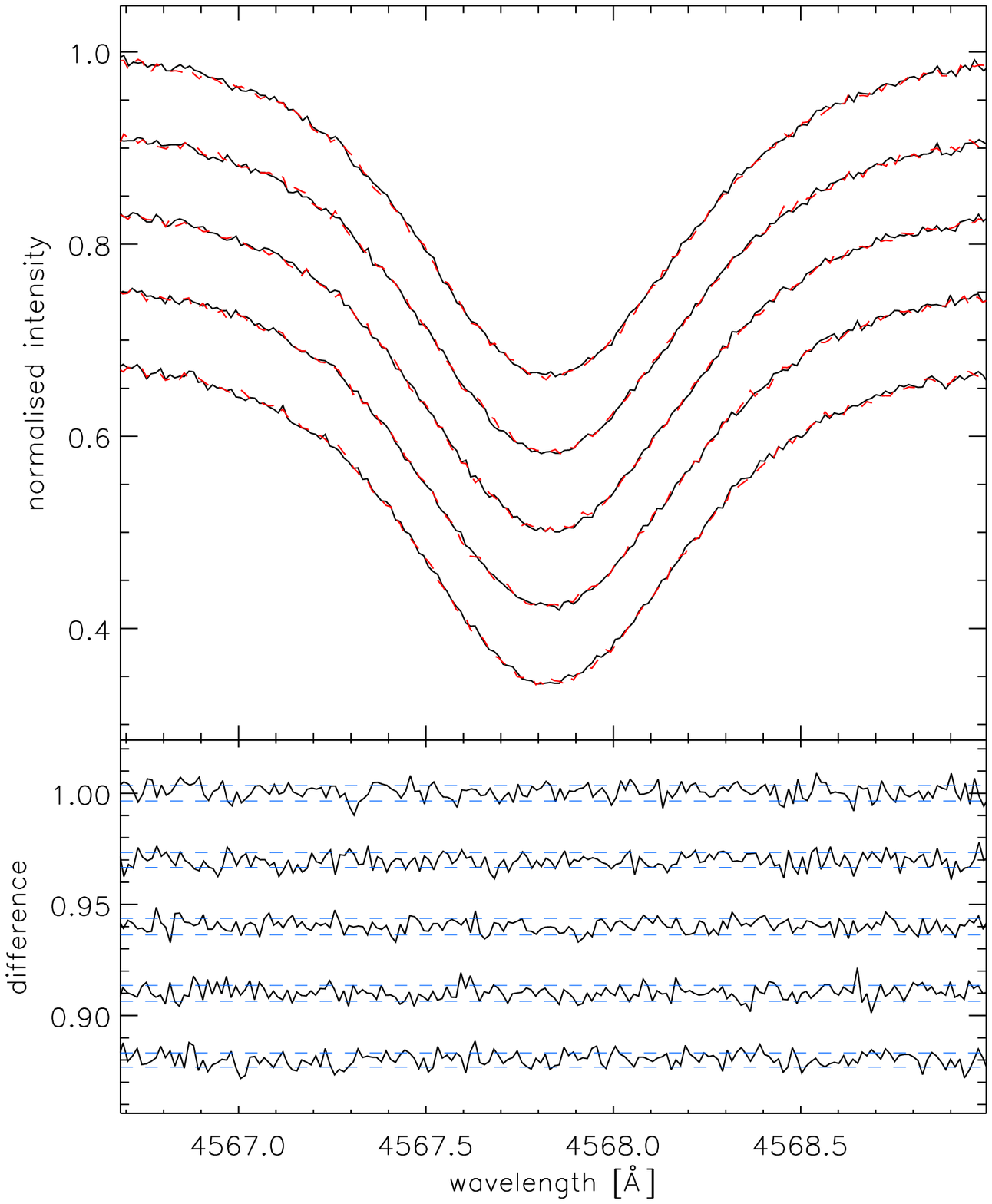}
\caption{Same as the plot on the right side of Fig.~\ref{fig:hd44743-lsd2}, but for $\epsilon$\,CMa.}
\label{fig:hd52089-profiles2}
\end{figure}
}
%--------------------------------------------------------------------

As expected, given the shorter exposure times compared to the December 2013 observations, the noise of the individual consecutive Stokes $V$ LSD profiles was in most cases too strong to lead to a field detection. By averaging the profiles, we obtained an extremely high S/N (see Table~\ref{tab:Bfield}) which led to a solid definite detection. For $\epsilon$\,CMa we derived an average longitudinal magnetic field strength of about $-$4\,G with uncertainties of 2--2.5\,G, depending on the adopted line mask.

\begin{figure}[]
\includegraphics[width=85mm,clip]{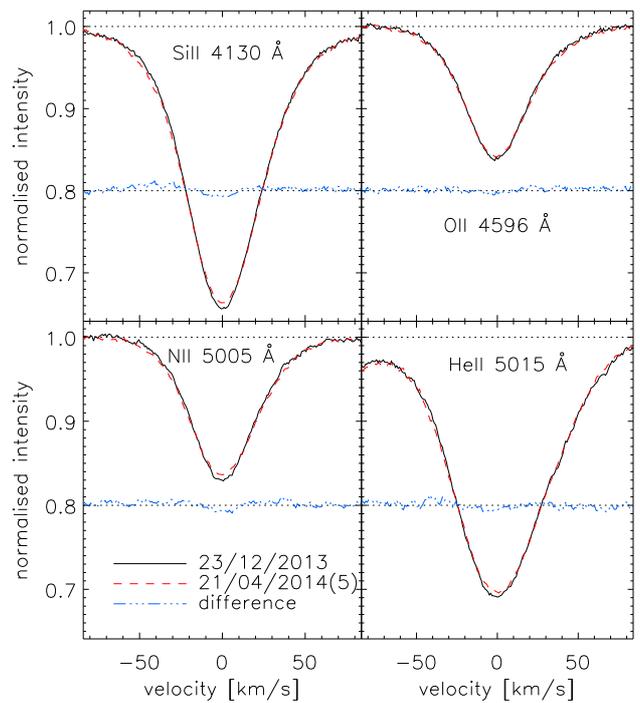}
\caption{Comparison between the line profiles of $\epsilon$\,CMa observed for the \ion{Si}{ii} 4130\,\AA\ (top left), \ion{O}{ii} 4596\,\AA\ (top right), \ion{N}{ii} 5005\,\AA\ (bottom left), and \ion{He}{ii} 5015\,\AA\ (bottom right) lines on the nights of 23 December 2013 and 21 April 2014, last exposure. Both profiles have been shifted using the radial velocity determined from the spectrum obtained on 23 December 2013. The blue dash-dotted line shows the difference between the two profiles, rigidly shifted upwards by 0.8.}
\label{fig:hd52089-lp-variations}
\end{figure}

In light of the magnetic field detection presented here, it is important to notice that the HARPS data show small line profile variations, in particular for silicon and nitrogen. Figure~\ref{fig:hd52089-lp-variations} shows a comparison between representative line profiles of four elements observed on 23 December 2013 and on 21 April 2014, last exposure. Variations of similar magnitude are also visible for the other lines of the elements shown in Fig.~\ref{fig:hd52089-lp-variations}. Line profile variations of similar amplitude are also visible using the other collected spectra, except when comparing only the spectra obtained on 21 April 2014, so that the period of the line profile variations should be much longer than one\,hour. Though small, the line profile variations shown in Fig.~\ref{fig:hd52089-lp-variations} seem to be significant: for example, the lines observed on 21 April 2014 appear to be systematically shallower that those obtained on 23 December 2013.
%--------------------------------------------------------------------
%--------------------------------------------------------------------

The line profile variations we detected might be caused by the presence of either surface spots or pulsation, since the star is located in the $\beta$\,Cep instability strip. The star is in the region of the sky and magnitude range covered by the BRITE satellites \citep{weiss2014}, whose data will have the precision needed to access the origin and periodicity of the observed line profile variations.
\subsubsection{Magnetic field geometry and strength}\label{geometry_and_strength_hd52089}
Also $\epsilon$\,CMa shows little variations of the Stokes $V$ LSD profiles. This is highlighted by the \bz\ values we obtained for this star appearing to be rather constant, with a small scatter around $-$5\,G (see Fig.~\ref{fig:bz}). For this star, we found a \vsini\ value of 21.2\,$\pm$\,2.2\,\kms\ that, combined with the stellar radius, leads to a maximum rotation period of about 24\,days. The star's critical velocity of 389\,\kms\ leads to a lower limit on the rotation period of 1.3\,days.

The magnetic field measurements conducted on the basis of FORS1 observations in 2006 and 2007 \citep[see Sect.\ref{introduction};][]{hubrig2009,bagnulo2012} also have to be taken into account to infer the star's geometry. Following the conclusions of \citet{bagnulo2012} and the terminology used for LSD profiles, the 5$\sigma$ detection obtained from the 2007 FORS1 data could be considered as a ``marginal detection''. Since the FORS1 results are at the limit of being considered significant and because of the known impact of the data reduction procedure on the \bz\ measurements \citep{bagnulo2012,bagnulo2013}, we re-reduced and analysed both FORS1 datasets using two different and independent routines and codes (Bonn and Potsdam).

For the first reduction (Bonn), we used a suite of IRAF\footnote{Image Reduction and Analysis Facility (IRAF -- \url{http://iraf.noao.edu/}) is distributed by the National Optical Astronomy Observatory, which is operated by the Association of Universities for Research in Astronomy (AURA) under cooperative agreement with the National Science Foundation.} \citep{tody} and IDL routines that follow the technique and the recipes presented by \citet{bagnulo2002,bagnulo2012}\footnote{More details about the applied data reduction and analysis procedure of FORS spectropolarimetric data will be given in a separate work; Fossati et al., in prep.}. The results obtained using the whole spectrum, as done by \citet{hubrig2009} and \citet{bagnulo2012} are listed in Table~\ref{tab:fors}.
%--------------------------------------------------------------------
\begin{table}[h!]
\caption[]{Summary of the results obtained from four different reductions of the FORS1 data of $\epsilon$\,CMa.}
\label{tab:fors}
\begin{center}                     
\begin{tabular}{lcc}
\hline
\hline
Analysis & 2006 & 2007 \\
 & \multicolumn{2}{c}{\bz\ [G]} \\
\hline
\citet{hubrig2009}  & $-$200\,$\pm$\,48 & $-$129\,$\pm$\,34 \\
\citet{bagnulo2012} & $-$127\,$\pm$\,60 & $-$196\,$\pm$\,37 \\
This work (Bonn)    & $-$138\,$\pm$\,40 & $-$149\,$\pm$\,32 \\
This work (Potsdam) & $-$212\,$\pm$\,42 & $-$133\,$\pm$\,43 \\
\hline
\end{tabular}
\end{center}                     
%\tablefoot{}
\end{table}

%--------------------------------------------------------------------

For the second reduction (Potsdam), we used the suite of tools described in \citet{steffen2014}. The results are given in Table~\ref{tab:fors}. Keeping in mind that different data reduction procedures can lead to different results \citep{bagnulo2012,bagnulo2013}, the values reported in Table~\ref{tab:fors} are in rough agreement. Despite this, there is a large difference between the magnetic field measured with FORS1 and HARPS. In this case, the discrepancy is significant because a purely dipolar magnetic field configuration with a rotation period ranging between 1.3 and 24\,days will hardly be able to fit both HARPS and FORS1 measurements. In this respect, one has to consider that systematic differences in the \bz\ measurements obtained with two different instruments are common, as reported by \citet{landstreet2014}. In this light, further measurements of the magnetic field of $\epsilon$\,CMa are clearly needed in order to constrain the magnetic field geometry and strength. To be conservative and to allow for different interpretations of the results, we only used the HARPS \bz\ measurements to derive the lower limit of the dipolar magnetic field strength, obtaining B$_{\mathrm d}\gtrapprox$13\,G.

In the context of the classification of massive star magnetospheres presented by \citet{petit2013} and assuming a minimum dipolar magnetic field strength of 13\,G and a maximum rotation period of 24\,days, we obtained a lower limit on the Alfv{\'e}n radius of about 1.8 stellar radii and an upper limit on the Keplerian corotation radius of about 7.9 stellar radii. The star is therefore likely to have a dynamical magnetosphere. 
\section{Discussion}\label{discussion}
\subsection{Mixing and nitrogen abundances}\label{mixing}
The signature of nitrogen enhancement in magnetic stars is not clear-cut. There are magnetic stars that are not enhanced and non-magnetic stars that show enhanced nitrogen \citep[see, e.g.,][]{morel2012}. In this respect, our two targets are not exceptional. While the nitrogen abundance of $\beta$\,CMa agrees
perfectly with the one derived from unmixed early B-type stars in the solar neighbourhood \citep{nieva2012}, that of $\epsilon$\,CMa appears slightly enhanced (0.2-0.3\,dex). In addition, the CNO abundances derived for both stars follow the nuclear path in the N/C vs. N/O diagram perfectly \citep{przybilla2010,langer2012,maeder2014}. 

Unfortunately, there is no clear implication from these results in terms of magnetic field origin. If the fields were fossils, some of the magnetic stars could have been born rapidly rotating, with the consequence of mixing up some nitrogen before spinning down \citep{meynet2011}. In addition, if the fields were formed in binary mergers, then according to the merger simulations of \citet{glebbeek2013}, a ubiquitous nitrogen enhancement is expected only for merger products that are more massive than 20$\,M_{\odot}$, while they find that lower mass objects may or may not show some nitrogen enhancement. 
\subsection{Lack of a ``magnetic desert'' in massive stars}\label{desert}
\begin{figure*}[]
\sidecaption
\includegraphics[width=129mm,clip]{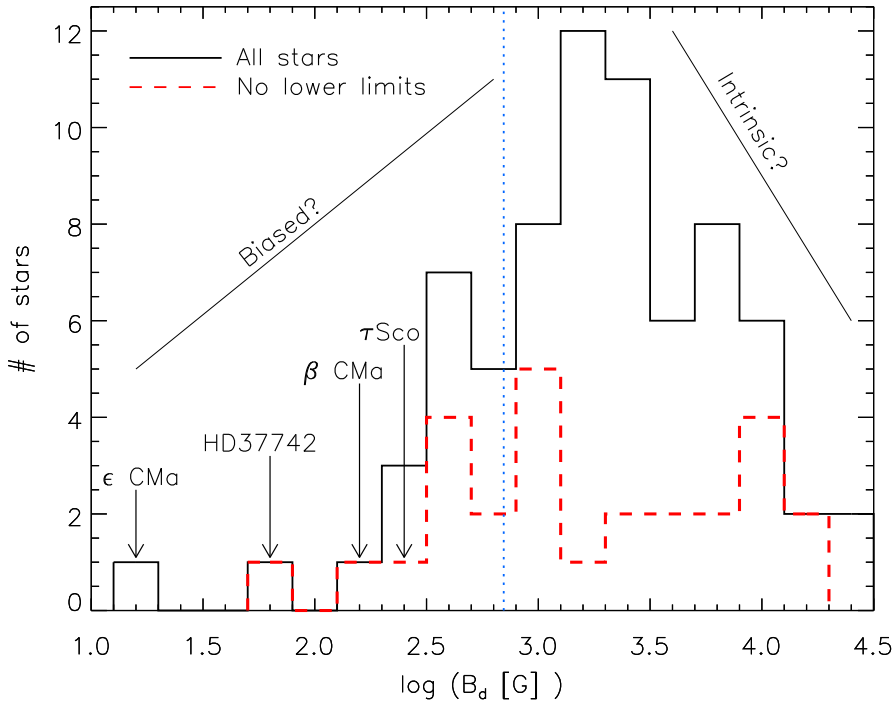}
\caption{Number of magnetic massive stars as a function of the logarithm of the dipolar magnetic field strength B$_{\mathrm d}$. The histogram was drawn using a fixed bin size of $\Delta$log(B$_{\mathrm d}$)\,=\,0.2\,dex. The histogram drawn with a dashed line was obtained considering only those stars for which a definite value of B$_{\mathrm d}$ is available (i.e., excluding the stars for which only a lower limit on B$_{\mathrm d}$ is available). The arrows show the position of the four stars with the weakest magnetic field (for $\epsilon$\,CMa the B$_{\mathrm d}$ value is a lower limit). The vertical dotted line indicates the magnetic field detection threshold of the FORS low-resolution spectropolarimeters (see text for more details).}
\label{fig:histogram}
\end{figure*}

To different degrees of certainty, the magnetic field of the two target stars appears to be weak. This is seen in a broader context in Fig.~\ref{fig:histogram}, which shows the number of magnetic massive stars as a function of the logarithm of their dipolar magnetic field strength. In addition to the two magnetic stars analysed here, the histogram includes the 
sample of stars presented by \citet{petit2013}, \citet{fossati2014}, and \citet{alecian2014}. We highlight the position of the four stars that present the weakest magnetic fields. The star $\epsilon$\,CMa seems to hold the weakest dipolar magnetic field strength, but the value we derived is only a lower limit. The star with the second weakest magnetic field, HD\,37742 ($\zeta$\,Ori\,Aa), was included in the study by \citet{petit2013}, though \citet{bouret2008} did not obtain a definite magnetic field detection. Nevertheless, subsequent \espa\ spectropolarimetric observations confirmed the presence of a weak magnetic field and also revealed that $\zeta$\,Ori\,A is a long-period binary system \citep{hummel2013} where only the primary star appears to be magnetic \citep{blazere2014}. With increasing magnetic field strength, we encounter $\beta$\,CMa, for which we adopted B$_{\mathrm d}$=99.5\,G (see Sect.~\ref{geometry_and_strength_hd44743}), and $\tau$\,Sco. Also $\zeta$\,Cas (B2IV, V\,=\,3.66\,mag), not included in Fig.~\ref{fig:histogram}, presents a weak magnetic field with a dipolar field strength below 100\,G (Briquet et al., in prep). These detections of weak fields might indicate that magnetic fields in massive stars could be more ubiquitous than derived from the current number statistics, in particular when one accounts for observational biases.  
%--------------------------------------------------------------------
%--------------------------------------------------------------------

As a matter of fact, the histograms shown in Fig.~\ref{fig:histogram} contain  biases that we qualitatively discuss here. The most important bias affecting the distributions shown in Fig.~\ref{fig:histogram} is that weak fields are more difficult to detect than strong fields. The use of LSD or similar techniques, on the basis of high-resolution spectropolarimetric data (e.g., from \espa, Narval, HARPSpol), implies that it will be more likely to detect a weak magnetic field for bright, low-\vsini\ stars. It is thus not surprising that the stars with the weakest fields, with the exception of $\zeta$\,Ori\,Aa, are visually bright slow rotators. The implication is that many more massive stars can be expected to host weak fields. We note that this is a markedly different situation than for the intermediate-mass stars, where stars seem to either have rather strong magnetic fields of B$_{\mathrm d}\gtrsim$300\,G (magnetic chemically peculiar ApBp stars) or fields below 1\,G \citep{auriere2007,lignieres2009,donati2009,petit2011,lignieres2014}.
In contrast to the situation determined for intermediate-mass stars, there is no clear evidence of a ``magnetic desert'' for massive stars, though we cannot exclude that a ``magnetic desert'' is indeed present, but less extended than for intermediate-mass stars. Deeper observations of massive stars are needed, in order to place firmer constraints on the extent of the ``magnetic desert'' in massive stars.

The use of a low-resolution spectropolarimeter (e.g., FORS) removes the bias towards slow rotators and alleviates the bias towards brighter stars somewhat, 
but it allows magnetic fields to be unambiguously detected in massive stars only for B$_{\mathrm d}\gtrsim$700\,G (Fig.~\ref{fig:histogram}), where we assumed an average uncertainty on \bz\ of 40\,G, a detection threshold of 5$\sigma$, and B$_{\mathrm d}\gtrsim$3.3\,$<$B$_z>_{max}$. This implies that the histograms may be essentially bias-free above, say, 1000\,G. This value corresponds roughly to the peak of the distribution, suggesting again that the strong drop in the number of magnetic stars with fields below 1000\,G might be at least partly due to incompleteness. 
  
On the basis of these considerations, it appears clear that the distribution below about 1\,kG shown in Fig.~\ref{fig:histogram} might be incomplete, while the distribution above 1\,kG is likely to represent the true distribution. As a consequence, the number of magnetic massive stars might possibly be quite large and may be comparable to the $\sim$\,30\% of slow rotators found in the LMC early B-type single stars by \citet{dufton2013}. To prove or disprove this conclusion, it will be necessary in future surveys to aim for higher S/N observations to decrease the uncertainties on the field measurements.
\section{Conclusion}\label{conclusion}
Within the context of the BOB collaboration, whose primary aim is to characterise the incidence of magnetic fields in slowly rotating massive stars, we obtained HARPSpol high-resolution spectropolarimetric observations of the early B-type stars $\beta$\,CMa (HD\,44743 - B1\,II/III) and $\epsilon$\,CMa (HD\,52089 - B1.5\,II) on two different runs, the first in December 2013 and the second in April 2014. For both stars, we repeatedly detected the signature of a weak ($<$30\,G in absolute value) longitudinal magnetic field.

We used a combination of FEROS and HARPS data to constrain the atmospheric parameters and chemical abundances. For $\beta$\,CMa we obtained \Teff\,=\,24700$\pm$300\,K and \logg\,=\,3.78$\pm$0.08, while for $\epsilon$\,CMa we obtained \Teff\,=\,22500$\pm$300\,K and \logg\,=\,3.40$\pm$0.08. These results agree well with those previously obtained by other authors. We also confirm the previous finding that, of the two stars, only $\epsilon$\,CMa presents a surface overabundance of nitrogen. For both stars, we also derived the stellar parameters on the basis of two sets of stellar evolutionary tracks \citep{brott2011,georgy2013}, finding that the two stars have a similar mass of about 12.5\,\M. We also concluded that both stars are most likely core hydrogen burning and that they have already spent more than two-thirds of their main sequence life.

For $\beta$\,CMa, we performed a mode identification of the data presented by \citet{mazumdar2006} and obtained an equatorial rotational velocity of $v_{\rm eq}$\,=\,30.6\,$\pm$\,0.9\,\kms\ and an inclination angle $i$\,=\,56.7\,$\pm$\,1.7$^{\circ}$. This led to a rotation period of 13.6\,$\pm$\,1.2\,days. On the basis of these results we attempted a preliminary fit of a perfectly dipolar magnetic field in order to derive the magnetic field geometry and strength. At the 1$\sigma$ level, we obtained an obliquity $\beta$ ranging between about 5$^{\circ}$ and 90$^{\circ}$, and a dipolar magnetic field strength B$_{\mathrm d}$ ranging between about 60 and 230\,G. We obtained a best fitting $\beta$ of $\beta$=22.3$^{\circ}$ and B$_{\mathrm d}$ of 96.9\,G. The derived B$_{\mathrm d}$ value is below what is typically found for other magnetic massive stars \citep[see, e.g.,][]{petit2013,alecian2014}. \citet{ignace2013} showed that the X-ray spectrum of $\beta$\,CMa is softer than that of the magnetic $\tau$\,Sco-analogue stars. This further strengthens the conclusion that $\beta$\,CMa hosts a weak magnetic field.

The \vsini\ value and stellar parameters we derived for $\epsilon$\,CMa imply that the rotation period ranges between 1.3 and 24\,days. In addition to the HARPS spectra, we re-analysed FORS1 low-resolution spectropolarimetric observations to confirm the previous magnetic field detection obtained from data collected in 2007 \citep{hubrig2009,bagnulo2012}, obtaining a ``marginal'' detection at 4--5$\sigma$. The possible presence of systematic differences between \bz\ measurements obtained using different instruments does not allow us to draw any firm conclusion about the magnetic field strength and geometry. In case such systematic difference was not present, a purely dipolar magnetic field geometry could hardly fit both HARPS and FORS1 measurements. To be conservative and to allow further interpretations of the results, we
only used the HARPS measurements to derive the minimum dipolar magnetic field strength, obtaining B$_{\mathrm d}\gtrsim$13\,G. 

Our results imply that both stars are expected to have a dynamical magnetosphere, so that the magnetic field is too weak to support a circumstellar disk. This is confirmed by the lack of H$\alpha$ emission in the spectra.

The distribution of the dipolar magnetic field strength for the magnetic massive stars known to date (Fig.~\ref{fig:histogram}) shows the presence of a non-negligible number of stars with a weak magnetic field, and therefore the lack of a clear ``magnetic desert'' as observed for intermediate-mass stars \citep[e.g.,][]{auriere2007}. Considerations of the biases involved in the detection of (weak) magnetic fields with the current available instrumentation and techniques have led us to the conclusion that the number of stars hosting a magnetic field might be greater than what is currently observed, possibly up to the $\sim$\,30\% of slow rotators found in the LMC early B-type single stars by \citet{dufton2013}. 

The work presented here, as well as what was done in the past, on the detection of weak magnetic fields shows that weak fields in massive stars are indeed detectable and that more work still has to be done in order to characterise the real incidence and evolution of magnetic fields in massive stars. 
\begin{acknowledgements}
The authors would like to thank the referee, John Landstreet, for useful comments that improved the manuscript. LF acknowledges financial support from the Alexander von Humboldt Foundation. TM acknowledges financial support from Belspo for contract PRODEX GAIA-DPAC. NL and AR acknowledge financial support from the DFG-CONICYT International Collaboration Project (DFG-06). SS-D acknowledges funding by the Spanish Ministry of Economy and Competitiviness under the grants AYA2010-21697-C05-05, AYA2012-39364-C02-01, and Severo Ochoa SEV-2011-0187, and by the Canary Islands Governments under grant PID2010119. LF is greatly in debt to Oleg Kochukhov for providing most of the tools used here for reducing and analysing the HARPSpol data. LF also thanks Gregg Wade, John Landstreet, and Coralie Neiner for fruitful discussions. The authors thank Andreas Irrgang, Ignacio Negueruela, Henk Spruit, Gautier Mathys, and Alexander Kholtygin for useful comments. SH and MS thank Thomas Szeifert for providing the pipeline for the FORS spectra extraction.
\end{acknowledgements}
%
%

%
%\Online
%
%
%
%
\end{document}